\documentclass[preprint,aps]{revtex4}
\usepackage{graphicx}
\usepackage{amssymb}
\newcommand{\be}{\begin{eqnarray}}
\newcommand{\ee}{\end{eqnarray}}
\usepackage{hyperref}
\usepackage{cancel} 
\newcommand{\nn}{\nonumber}

\usepackage{amsmath,amssymb}

\begin{document}

\title{Wigner Distributions for Gluons in Light-front
Dressed Quark Model}

\author{\bf Asmita Mukherjee, Sreeraj Nair and Vikash Kumar Ojha}

\affiliation{ Department of Physics,
Indian Institute of Technology Bombay,\\ Powai, Mumbai 400076,
India.}
\date{\today}

\begin{abstract}
\noindent
We present a calculation of Wigner distributions for gluons in light-front
dressed quark model. We calculate the kinetic and canonical gluon orbital 
angular momentum and spin-orbit correlation of the gluons in this model. 

\end{abstract}

\maketitle

\section{Introduction}
\noindent
To have a complete understanding of matter at subatomic level, it is important to understand the
nucleon spin structure. This means an understanding of how the spin ($\frac{1}{2}$) of the 
nucleon is shared by the quarks and gluons in the nucleon and what is the contribution of
their orbital angular momentum (OAM). Earlier it was believed that all nucleon spin is carried
by quarks. EMC experiment showed that contribution of quark spin to nucleon spin is very small.
So, an important question is from where does the missing (remaining) angular momentum comes
from? As the nucleon is made up of quarks and gluons, it is natural to expect that the missing 
angular momentum come either from gluon spin or from quark or gluon OAM. This is expressed by spin 
sum rule \cite{jm}

\be\nn
\frac{1}{2}=\frac{1}{2}\sum_{q}{\Delta q}+\underbrace{\sum_{q}\mathcal{L}^q}_{\text{quark OAM}}
 + \Delta G + \underbrace{{\mathcal{L}^g}}_{\text{Gluon OAM}}
\ee
\noindent
Here $\frac{1}{2}\sum_{q}{\Delta q}$ and $\Delta G$ are quark and gluon spin angular momentum
respectively. The above sum rule is called canonical spin sum rule. Except quark intrinsic part, 
other terms depend upon specific gauge choice.

\noindent
Recently, Chen et al \cite{chen} introduced a gauge invariant extension (GIE) which is basically a 
prescription to find a manifestly gauge invariant quantity that coincides with a gauge 
non-invariant quantity in a particular gauge. In this way, one can extend the validity of a
physical interpretation of the canonical OAM $\mathcal{L}^q$ and $\mathcal{L}^g$ in light front
gauge to any other gauge. There is another decomposition called kinetic decomposition of nucleon
spin \cite{ji1}
\be\nn
\frac{1}{2}= S^q + L^q +J^g
\ee
where $S^q$ and $L^q$ are quark spin and kinetic orbital angular momentum
respectively. $J^g$ is total gluon
contribution to the angular momentum of nucleon. Later, Wakamatsu \cite{wak} further separated the $J^g$
into a gluon orbital part ($L^g$) and an intrinsic part ($S^g$) using a prescription similar
to \cite{chen}. Which of the two definitions of the OAM is  "physical" is a matter of intense debate. According
to the current understanding the difference between kinetic and canonical OAM is the choice of 
the Wilson line in the definition of non-local quark operator: a staple type gauge link gives 
canonical OAM whereas a straight line gauge gives kinetic OAM. Interesting physical interpretation
of both type of OAM is given in \cite{OAM_rev,mb}

\noindent
Recently it has been shown that the Wigner distribution can provide useful information on the spin and angular momentum 
correlation of quarks and gluons in the nucleon. Wigner distributions \cite{wigner} are quasi-probabilistic 
distribution
in which both position and momentum space information are encoded. They  are 
directly related to generalized parton correlation functions (GPCFs) \cite{metz}. GPCFs are fully
unintegrated off-diagonal quark-quark correlator and contain maximum amount of information about
the structure of nucleon. If we integrate GPCFs over light cone energy ($k^-$), we get generalized
transverse momentum dependent parton distribution functions (GTMDs). These GTMDs are Fourier 
transform of Wigner distributions and vice versa.

In fact, the GTMDs are directly related to generalized parton distributions (GPDs)
\cite{ji1} and
transverse
momentum dependent parton distribution functions (TMDs) \cite{metz}, both of which have been found to give
very useful information about the structure and spin of the nucleon in terms of quarks and gluons.
We can take two-dimensional Fourier transform of GPDs with respect to the momentum transfer in the
transverse direction to get impact parameter dependent parton distribution functions 
(IPDPDFs) \cite{burkardt}.
These give the correlation in transverse position and longitudinal momentum for different quark
and target polarization. On the other hand, TMDs provide momentum space information, and also the
strength of spin-orbit and spin-spin correlation. It has been shown that both of these 
distributions are linked to GTMDs. So we can consider the GTMDs, or equivalently Wigner 
distribution as mother distributions.

Wigner distributions described above, are joint position and momentum space distributions of 
quarks and gluons in the nucleon. Due to uncertainty principle, they are not positive definite 
and do not have probabilistic interpretation. However, integration of Wigner distributions over
one or more variables relates them to measurable quantities \cite{lorce}. Here model calculation of Wigner functions
are important as these calculations play a significant role in revealing what kind of information 
they can provide about the correlations of quarks and gluons inside the nucleon. Also such 
calculations are useful to check various relations among the GTMDs and the TMDs and GPDs. In fact
some relations have been found to hold only in certain class of models \cite{metz1}. 
In this work, we calculate the Wigner distributions in light front Hamiltonian approach \cite{pedestrian}. 
This approach gives a intuitive picture of deep inelastic scattering processes, as it is based
on field theory but keeps close contact with parton ideas \cite{kundu1}. Here the partons i.e quarks and gluons
are non-collinear, massive and they also interact. The target state is expanded in Fock space in
terms of multi-parton light front wave functions (LFWFs). The advantage of using light front 
(infinite momentum frame) formalism is that such wave functions are boost invariant, so one can 
work with a finite number of constituents of the nucleon and this picture is invariant under
Lorentz boost \cite{brod1}. In order to obtain the LFWFs of the nucleon, one needs a model 
light front Hamiltonian.
However many useful information can be obtained if one replaces the bound state by a simple
spin-$\frac{1}{2}$ composite relativistic state like a quark at one loop dressed with a gluon
\cite{hadron_optics1,hadron_optics2}.
In our previous work \cite{our1}, we studied 
the Wigner distribution of quarks for a simple relativistic spin-$\frac{1}{2}$ composite system,
namely for a quark dressed with a gluon, using light front Hamiltonian perturbation theory.
Here we calculate the Wigner distribution for gluons in the same model. This
calculation is useful as in most phenomenological models commonly
studied, gluonic degrees of freedom are not present \cite{lorce}, and a study of the
gluon spin and OAM is not possible there.  

\section{Wigner Distributions}

\noindent
The Wigner distribution for gluons can be defined as \cite{Jifeng}

\be
x W^{g}(x,\vec{k}_{\perp},\vec{b}_{\perp})=\int \frac{d^2 \vec{\Delta}_{\perp}}{(2\pi)^2}e^{-i\vec{\Delta}_{\perp}.\vec{b}_{\perp}} \int 
\frac{dz^{-}d^{2} z_{\perp}}{2(2\pi)^3 p^+}e^{i k.z}
 \nn \\ \Big{\langle } p^{+},\frac{\vec{\Delta}_{\perp}}{2},\sigma \Big{|}
\Gamma^{ij} F^{+i}\Big( -\frac{z}{2}\Big) F^{+j}\Big( \frac{z}{2}\Big) \Big{|}
p^{+},-\frac{\vec{\Delta}_{\perp}}{2},\sigma  \Big{\rangle }  \Big{|}_{z^{+}=0};
\label{eq1} \ee

\noindent
where  $\vec{\Delta}_{\perp}$ is the transverse momentum transfer from the target state and 
$\vec{b}_{\perp}$ is 2 dimensional vector
 in impact parameter space conjugate to $\vec{\Delta}_{\perp}$.
 We calculate Eq.~(\ref{eq1}) for $\Gamma^{ij} = \delta^{ij}$ and $\Gamma^{ij} = -i\epsilon^{ij}_{\perp}$.

\noindent
We have,
\be
F^{+i} = \partial^+ A^i -  \partial^i  A^+ + g f^{abc} A^{+}A^{i}.
\nn \ee

\noindent
The gauge field can be written as \cite{kundu1}, for $i=1,2$, 

\be
A^{i}\Big(\frac{z}{2}\Big) = 
\sum _{\lambda} \int \frac{dk^{+}d^{2} k_{\perp}}{2k^{+}(2\pi)^3}  
\Big[  \epsilon_{\lambda}^{i}(k) a_{\lambda}(k)e^{-\frac{i}{2}k.z} +   
\epsilon_{\lambda}^{*i}(k) a^{\dagger}_{\lambda}(k)e^{\frac{i}{2}k.z} \Big].
\nn \ee

\noindent
We choose the light front gauge, $A^ + = 0$, and take the gauge link to be
unity. 

\noindent
In our previous work \cite{our1} we calculated the quark Wigner distributions 
for a quark state dressed with a gluon. In this work we investigate  the gluon Wigner distribution 
in the same model using light-front Hamiltonian perturbation theory.  The state can be expanded in 
Fock space in terms of
multi-parton light-front wave functions (LFWFs) as \cite{hari} 
\be \label{dqs}
  \Big{| }p^{+},p_{\perp},\sigma  \Big{\rangle} = \Phi^{\sigma}(p) b^{\dagger}_{\sigma}(p)
 | 0 \rangle +
 \sum_{\sigma_1 \sigma_2} \int [dp_1]
 \int [dp_2] \sqrt{16 \pi^3 p^{+}}
 \delta^3(p-p_1-p_2) \nn \\ \Phi^{\sigma}_{\sigma_1 \sigma_2}(p;p_1,p_2) 
b^{\dagger}_{\sigma_1}(p_1) 
 a^{\dagger}_{\sigma_2}(p_2)  | 0 \rangle;
\ee

\noindent
where $[dp] =  \frac{dp^{+}d^{2}p_{\perp}}{ \sqrt{16 \pi^3 p^{+}}}$ and
$\sigma_1$ and $\sigma_2$ are the helicities of the quark and gluon respectively. 
The LFWFs ($ \Phi^{\sigma}(p)$ and $ \Phi^{\sigma}_{\sigma_1 \sigma_2}$) appearing in Eq.~(\ref{dqs}) are calculated by solving the light-front eigenvalue equation
in the Hamiltonian approach.
$ \Phi^{\sigma}(p)$ is the single particle  (quark) LFWF gives  the wave function re-normalization for the quark and $ \Phi^{\sigma}_{\sigma_1 \sigma_2}$ is the 
 two particle (quark-gluon) LFWF.
$ \Phi^{\sigma}_{\sigma_1 \sigma_2}(p;p_1,p_2)$ gives the probability
amplitude to find a bare 
quark having momentum $p_1$ and helicity $\sigma_1$ and a bare gluon with momentum $p_2$ and 
helicity $\sigma_2$ in the dressed quark. The two particle LFWF 
is related to the boost invariant LFWF; $\Psi^{\sigma}_{\sigma_1
\sigma_2}(x, q_\perp) =   
\Phi^{\sigma}_{\sigma_1 \sigma_2}
\sqrt{p^+}$.  Here we have used the Jacobi momenta $(x_i, q_{i \perp})$ : 
\be 
p_i^+= x_i p^+, ~~~~~~~~~~q_{i \perp}= k_{i \perp}+x_i p_\perp;
\ee
so that $\sum_i x_i=1, \sum_i q_{i\perp}=0$.   
These two-particle LFWFs can be calculated perturbatively as \cite{hari}:
\be \label{tpa}
\Psi^{\sigma a}_{\sigma_1 \sigma_2}(x,q_{\perp}) = 
\frac{1}{\Big[    m^2 - \frac{m^2 + (q_{\perp})^2 }{x} - \frac{(q_{\perp})^2}{1-x} \Big]}
\frac{g}{\sqrt{2(2\pi)^3}} T^a \chi^{\dagger}_{\sigma_1} \frac{1}{\sqrt{1-x}}
\nn \\ \Big[ 
-2\frac{q_{\perp}}{1-x}   -  \frac{(\sigma_{\perp}.q_{\perp})\sigma_{\perp}}{x}
+\frac{im\sigma_{\perp}(1-x)}{x}\Big]
\chi_\sigma (\epsilon_{\perp \sigma_2})^{*}.
\ee

\noindent
We use the  two component formalism \cite{zhang}. $\chi$ is the two
component fermion spinor.  $T^a$ are the usual color $SU(3)$ matrices, $m$ is 
the mass of the quark and $\epsilon_{\perp \sigma_2}$ is the polarization
vector of the gluon. \\

\noindent
The gluon-gluon correlator in Eq.~(\ref{eq1}) for a quark state dressed with
a gluon can be expressed in terms of the overlap of two-particle LFWFs. The
single particle sector of the Fock space expansion contributes only at $x=1$
and we exclude this. 

\noindent 
For $\Gamma^{ij} = \delta^{ij}$ we get
\be
W^{\sigma \sigma'}_{1}(x,k_{\perp},b_{\perp}) = -\sum_{\sigma_1, \sigma_{2},\lambda_{1}}  \int \frac{d^2 \Delta_{\perp}}{2(2\pi)^2}e^{-i\Delta_{\perp}.b_{\perp}} 
\Bigg[\Psi^{*\sigma'}_{\sigma_1 \lambda_{1} }(\hat{x},\hat{q}'_{\perp}) \Psi^{\sigma}_{\sigma_1 \sigma_2}(\hat{x},\hat{q}_{\perp})
\Big(\epsilon_{\sigma_2}^{1} \epsilon_{\lambda_1}^{*1}+\epsilon_{\sigma_2}^{2} \epsilon_{\lambda_1}^{*2}\Big)\Bigg];
\ee

\noindent
and for $\Gamma^{ij} = -i\epsilon^{ij}_{\perp}$ we get
\be
W^{\sigma \sigma'}_{2}(x,k_{\perp},b_{\perp}) = -i\sum_{\sigma_1, \sigma_{2},\lambda_{1}}  \int \frac{d^2 \Delta_{\perp}}{2(2\pi)^2}e^{-i\Delta_{\perp}.b_{\perp}} 
\Bigg[ \Psi^{*\sigma'}_{\sigma_1 \lambda_{1} }(\hat{x},\hat{q}'_{\perp}) \Psi^{\sigma}_{\sigma_1 \sigma_2}(\hat{x},\hat{q}_{\perp})
\Big(\epsilon_{\sigma_2}^{1} \epsilon_{\lambda_1}^{*2}-\epsilon_{\sigma_2}^{2} \epsilon_{\lambda_1}^{*1}\Big)\Bigg];
\ee

\noindent
where $ \hat{x} = (1-x)$ and $\hat{q}_{\perp}=-q_{\perp}$. Jacobi relation for the 
transverse momenta in the symmetric frame is given by 
$q'_{\perp} = k_{\perp}-\frac{\Delta_{\perp}}{2}(1-x)$ and $q_{\perp} 
= k_{\perp}+\frac{\Delta_{\perp}}{2}(1-x)$.
We represent the gluon Wigner distribution as $W^{\lambda\lambda^\prime}$, 
where $\lambda$ and $\lambda^\prime$ are polarization of target state and gluon respectively.\\

\noindent
We consider only longitudinally polarized target state and then we have four gluon Wigner distributions as follows, in
a manner similar to quark Wigner distributions \cite{lorce}

\noindent
Wigner distribution of unpolarized gluon in unpolarized target state as
\be
W^{UU}=W^{\uparrow \uparrow }_{1}(x,k_{\perp},b_{\perp})+ W^{\downarrow \downarrow }_{1}(x,k_{\perp},b_{\perp}); 
\ee
\noindent
Wigner distribution corresponding to the distortion due to longitudinal polarization of the target as 
\be
W^{LU}=W^{\uparrow \uparrow }_{1}(x,k_{\perp},b_{\perp})  -W^{\downarrow \downarrow }_{1}(x,k_{\perp},b_{\perp});
\ee 
\noindent
Wigner distribution corresponding to the distortion due to longitudinal polarization of the gluons as 
\be
W^{UL}=W^{\uparrow \uparrow }_{2}(x,k_{\perp},b_{\perp}) +W^{\downarrow \downarrow }_{2}(x,k_{\perp},b_{\perp}); 
\ee 
\noindent
and the Wigner distribution corresponding to the correlation due to longitudinal polarization 
of the target state and the gluons 
\be
W^{LL}=W^{\uparrow \uparrow }_{2}(x,k_{\perp},b_{\perp})  -W^{\downarrow \downarrow }_{2}(x,k_{\perp},b_{\perp}). 
\ee 
The final expressions for these four gluon Wigner distributions are given
by, using the two-particle LFWFs \\
\be
\label{rhouu}
W^{UU} (x,k_{\perp},b_{\perp}) = 
N \int d \Delta_x \int d \Delta_y  
\frac{ \mathrm{ cos}(\Delta_\perp\cdot b_\perp)}{D(q_{\perp})D(q'_{\perp})}
\Big[\frac{-4\Big((q_{\perp}q'_{\perp})(x^2-2x+2)+m^2 x^4\Big)}{x^3(x-1)^2}\Big];
\ee
\be
\label{rholu}
W^{LU} (x,k_{\perp},b_{\perp}) = 
N \int d \Delta_x \int d \Delta_y  
\frac{ \mathrm{ sin}(\Delta_\perp\cdot b_\perp)}{D(q_{\perp})D(q'_{\perp})}
\Big[\frac{4(2-x)(q_2 q'_1 - q_1 q'_2)}{x^2(x-1)^2}\Big];
\ee \\
\be
\label{rhoul}
W^{UL} (x,k_{\perp},b_{\perp}) = 
N \int d \Delta_x \int d \Delta_y  
\frac{ \mathrm{ sin}(\Delta_\perp\cdot b_\perp)}{D(q_{\perp})D(q'_{\perp})}
\Big[\frac{4(x^2-2x+2)(q_2 q'_1 - q_1 q'_2)}{x^3(x-1)^2}\Big];
\ee \\
\be
\label{rholl}
W^{LL} (x,k_{\perp},b_{\perp}) = 
N \int d \Delta_x \int d \Delta_y  
\frac{ \mathrm{cos}(\Delta_\perp\cdot b_\perp)}{D(q_{\perp})D(q'_{\perp})}
\Big[\frac{4\Big((q_{\perp}q'_{\perp})(2-x)+ m^2 x^3\Big)}{x^2(x-1)^2}\Big];
\ee
where $A_x ,A_y$ are $x,y$ component of $A_\perp$ and
\be
D(k_{\perp})=   \Big(m^2 - \frac{m^2 + (k_{\perp})^2 }{1-x} - \frac{(k_{\perp})^2}{x}\Big)
\hspace{0.5cm},\hspace{0.5cm}N=\frac{g^2}{2(2\pi)^2}.
 \nn \ee

\begin{figure}[!htp]
\begin{minipage}[c]{1\textwidth}
\tiny{(a)}\includegraphics[width=7.8cm,height=6cm,clip]{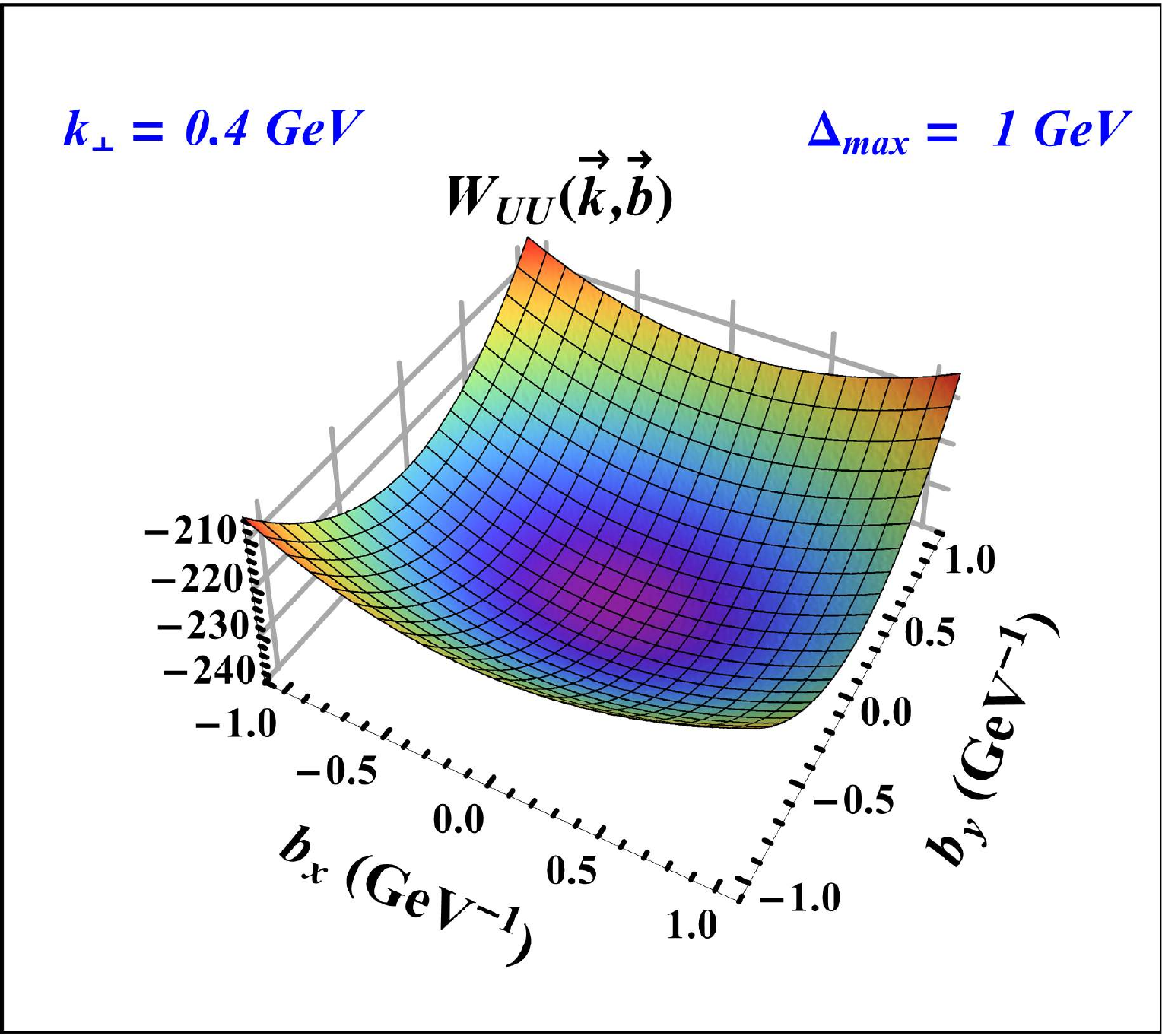}
\hspace{0.1cm}
\tiny{(b)}\includegraphics[width=7.8cm,height=6cm,clip]{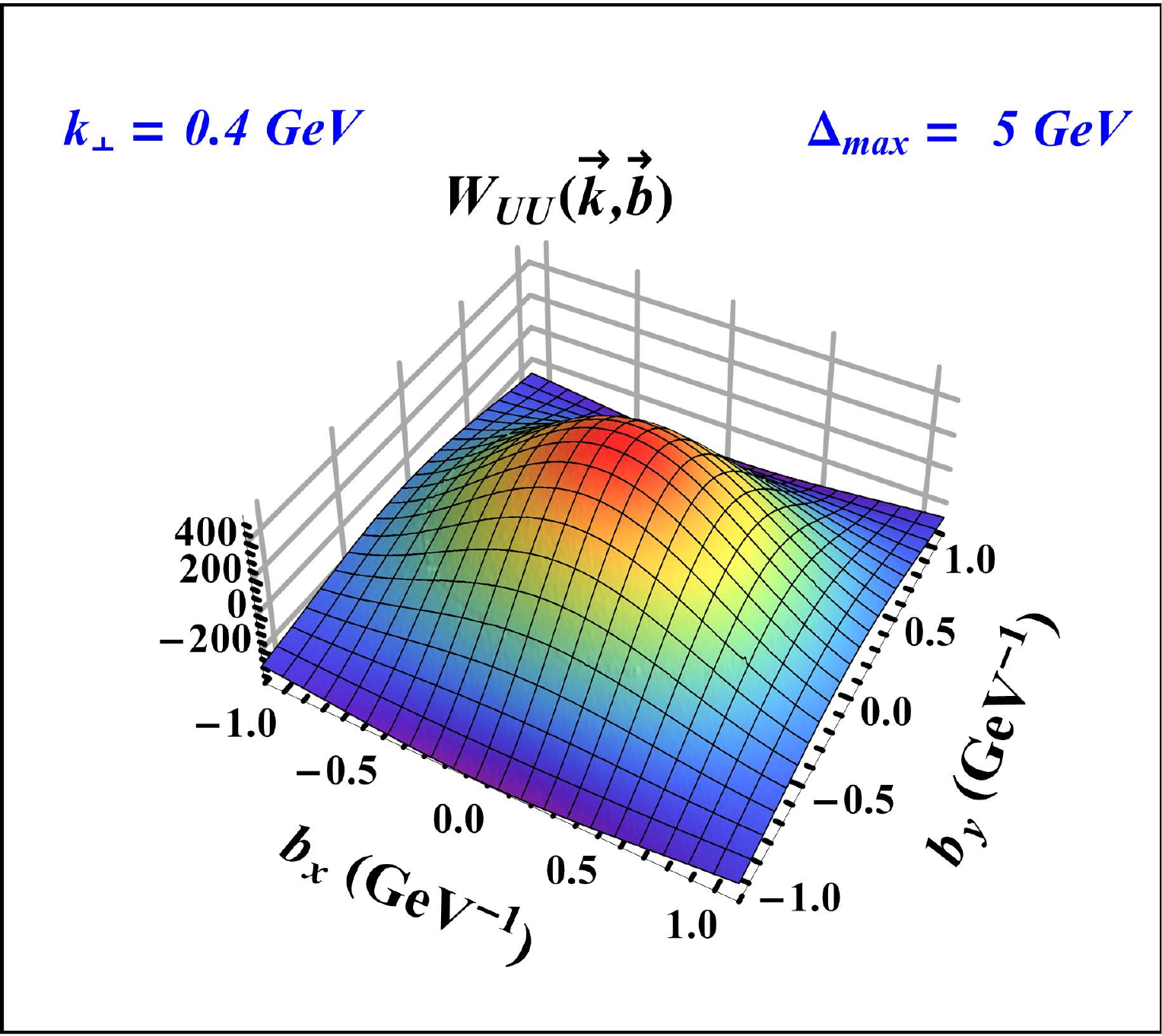}
\end{minipage}
\begin{minipage}[c]{1\textwidth}
\tiny{(c)}\includegraphics[width=7.8cm,height=6cm,clip]{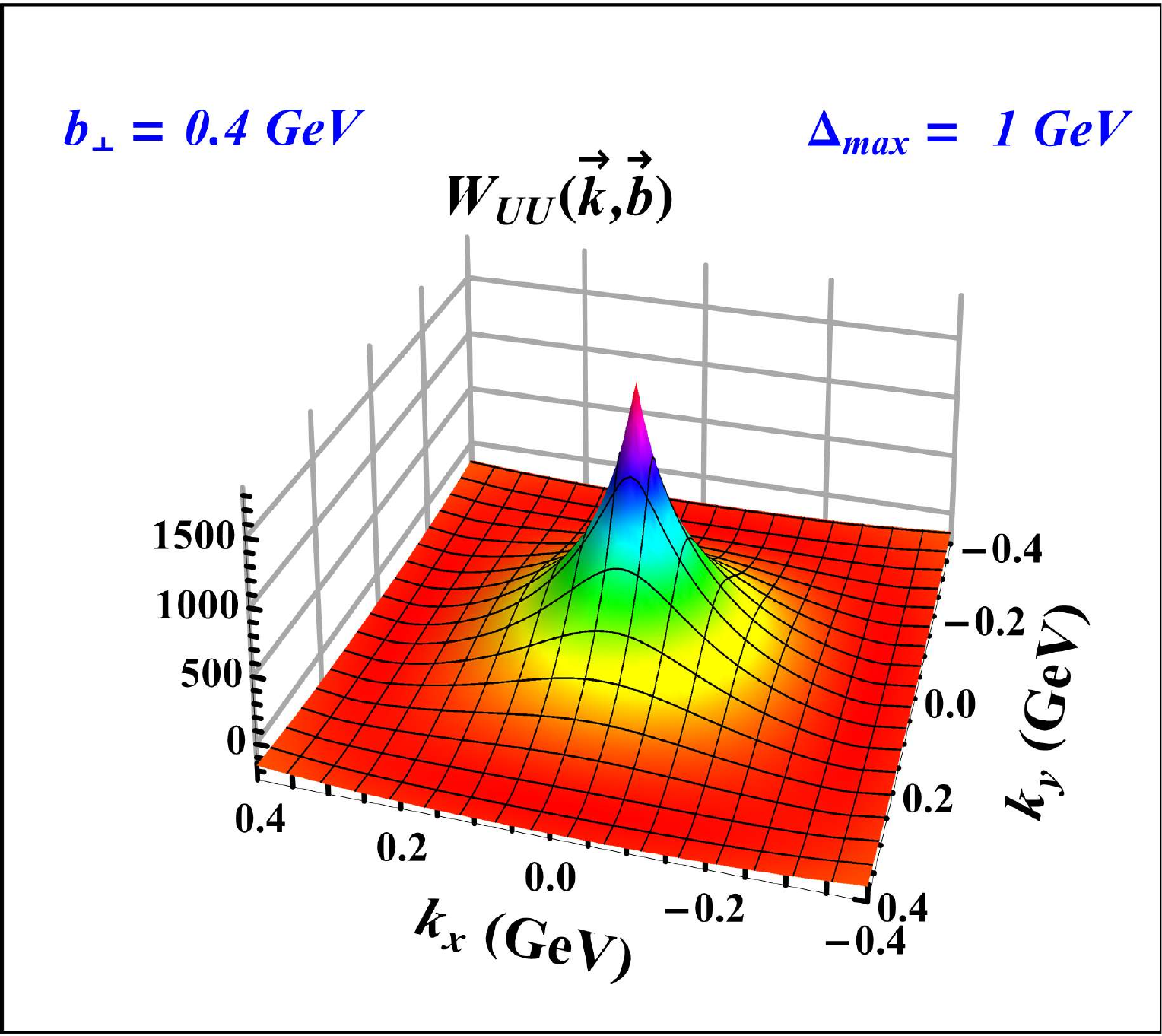}
\hspace{0.1cm}
\tiny{(d)}\includegraphics[width=7.8cm,height=6cm,clip]{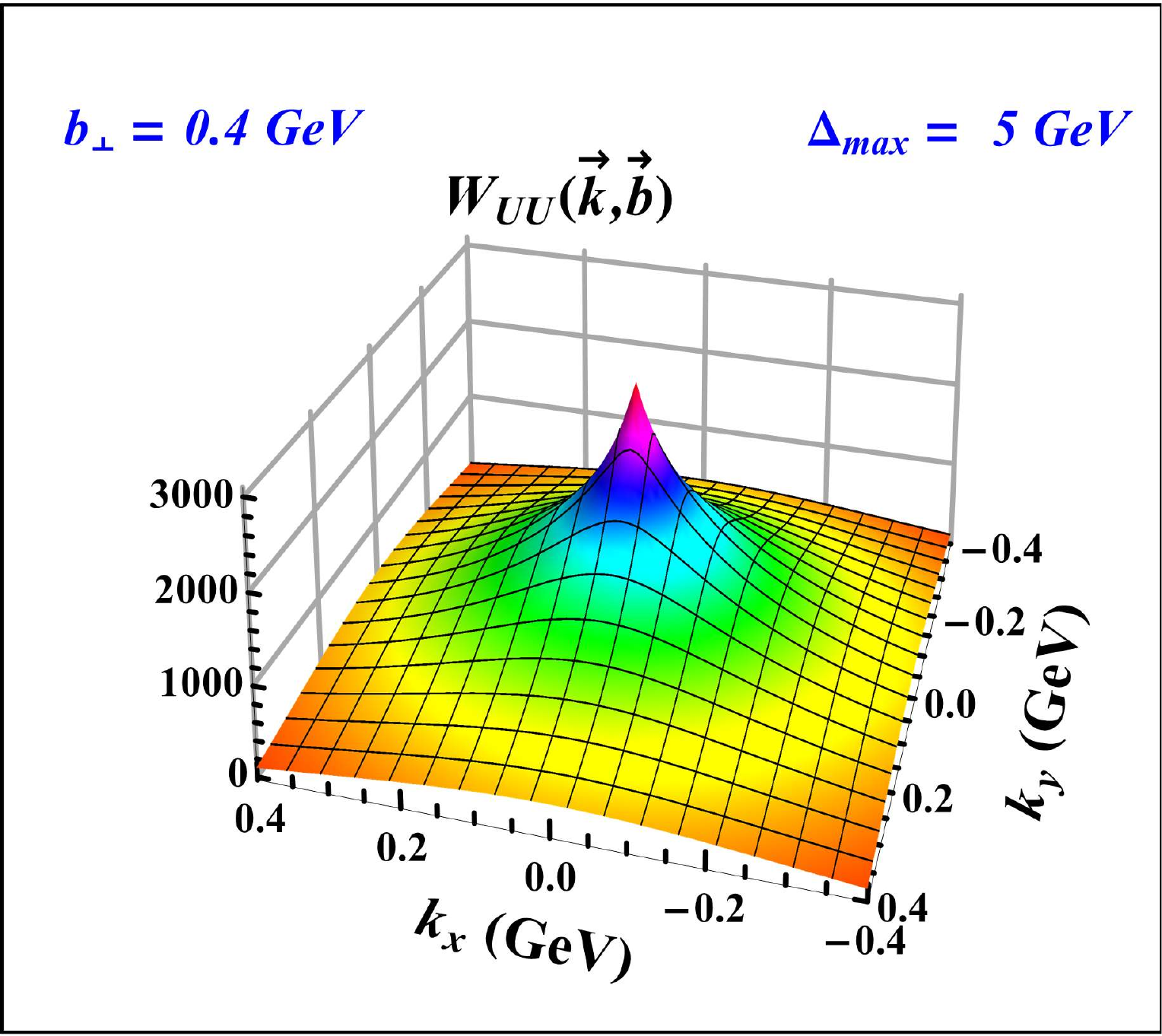}
\end{minipage}
\begin{minipage}[c]{1\textwidth}
\tiny{(e)}\includegraphics[width=7.8cm,height=6cm,clip]{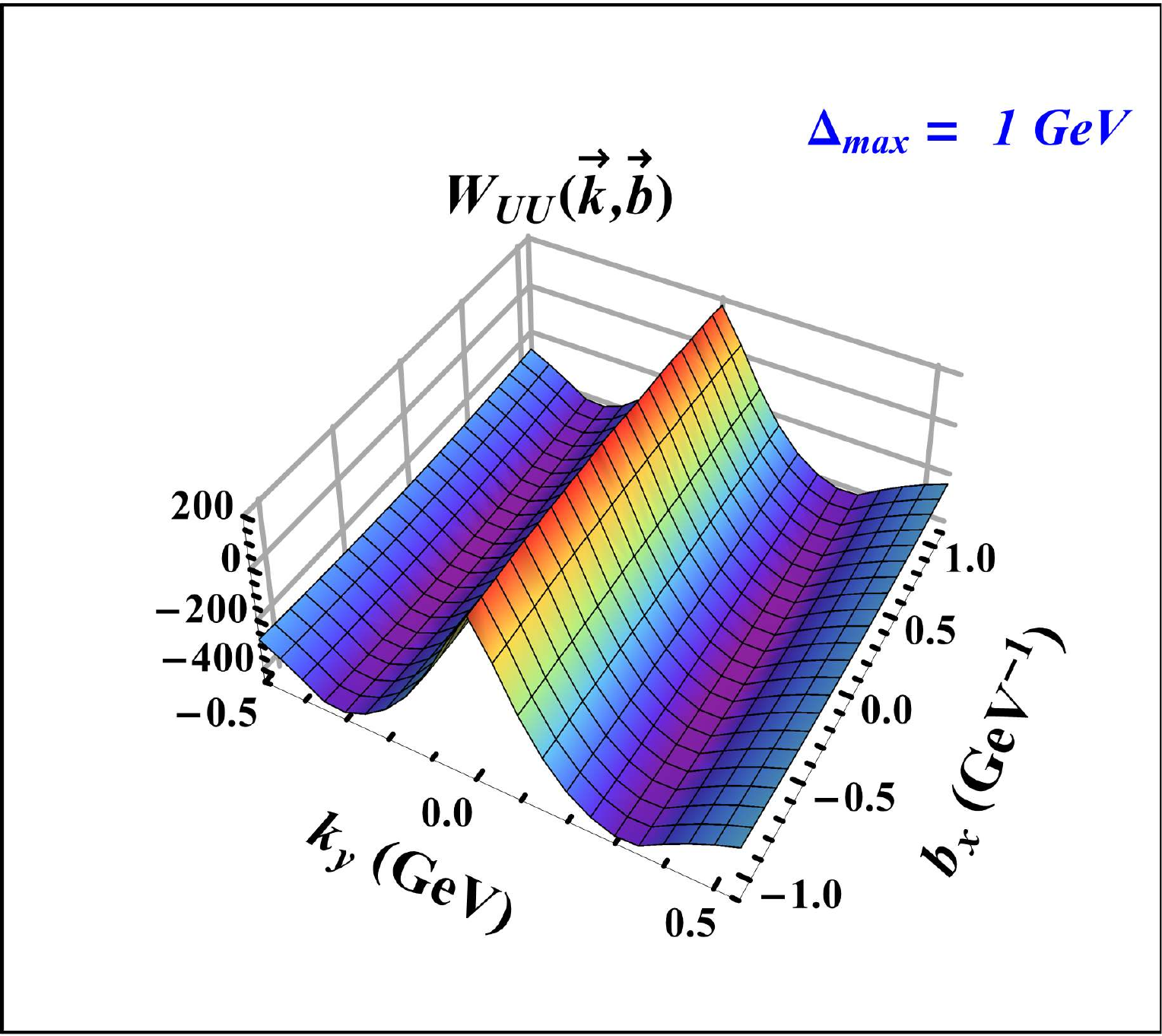}
\hspace{0.1cm}
\tiny{(f)}\includegraphics[width=7.8cm,height=6cm,clip]{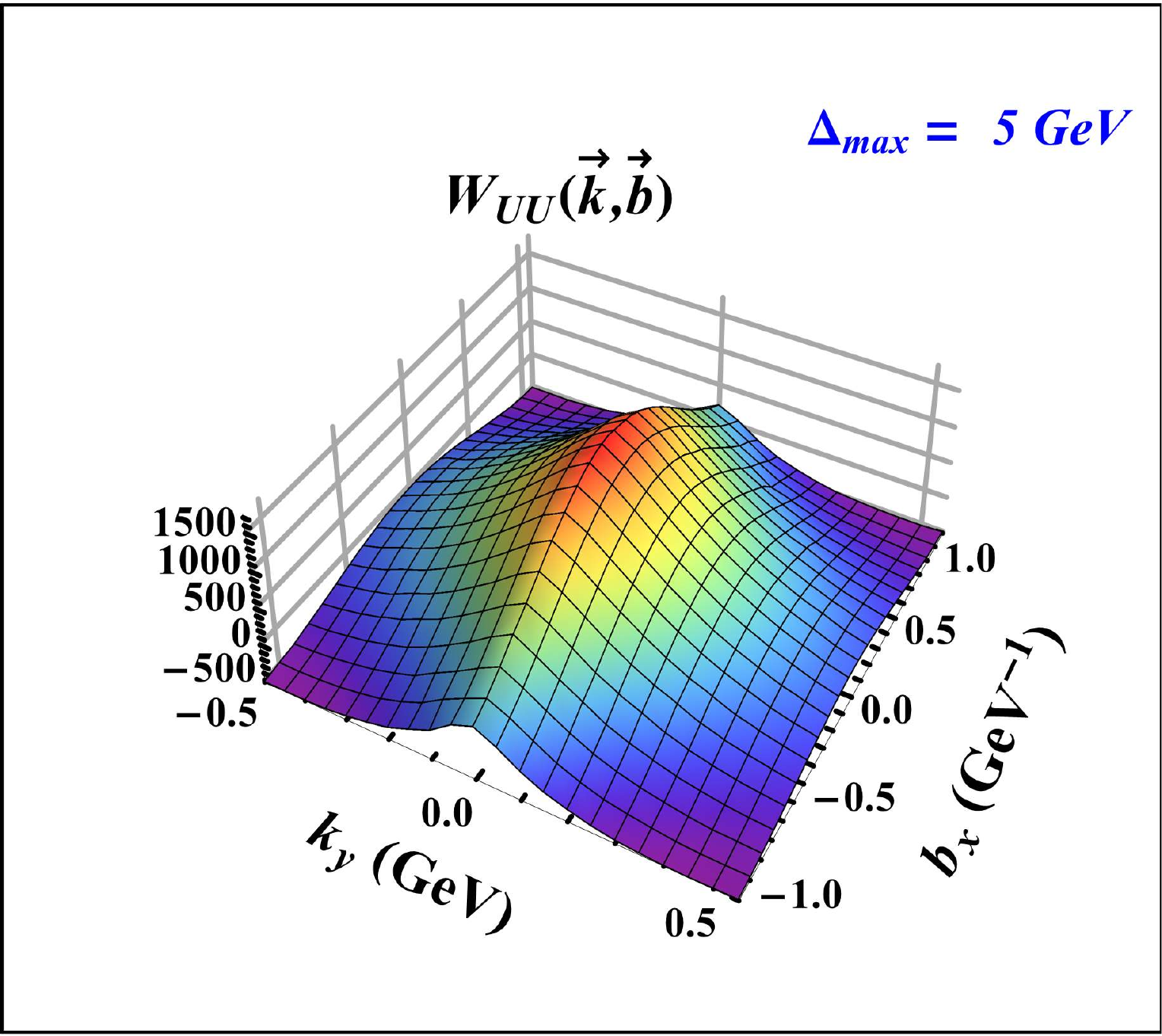}
\end{minipage}
\caption{\label{fig1}(Color online)
3D plots of the Wigner distributions $W^{UU}$. Plots (a) and (b) are in $b$ space with $k_\perp = 0.4$ GeV.
Plots (c) and (d) are in $k$ space with $b_\perp = 0.4$ $ \mathrm{GeV}^{-1}$.
Plots (e) and (f) are in mixed space where $k_x$ and $b_y$ are integrated.
All the plots on the left panel (a,c,e) are for $\Delta_{max} = 1.0$ GeV. Plots on the right panel (b,d,f) are for $\Delta_{max} = 5.0$ GeV.
For all the plots we kept $m = 0.33$ GeV, integrated out the $x$ variable and we took $\vec{k_\perp} = k \hat{j}$ 
and $\vec{b_\perp} = b \hat{j}$.  }
\end{figure}

\begin{figure}[!htp]
\begin{minipage}[c]{1\textwidth}
\tiny{(a)}\includegraphics[width=7.8cm,height=6cm,clip]{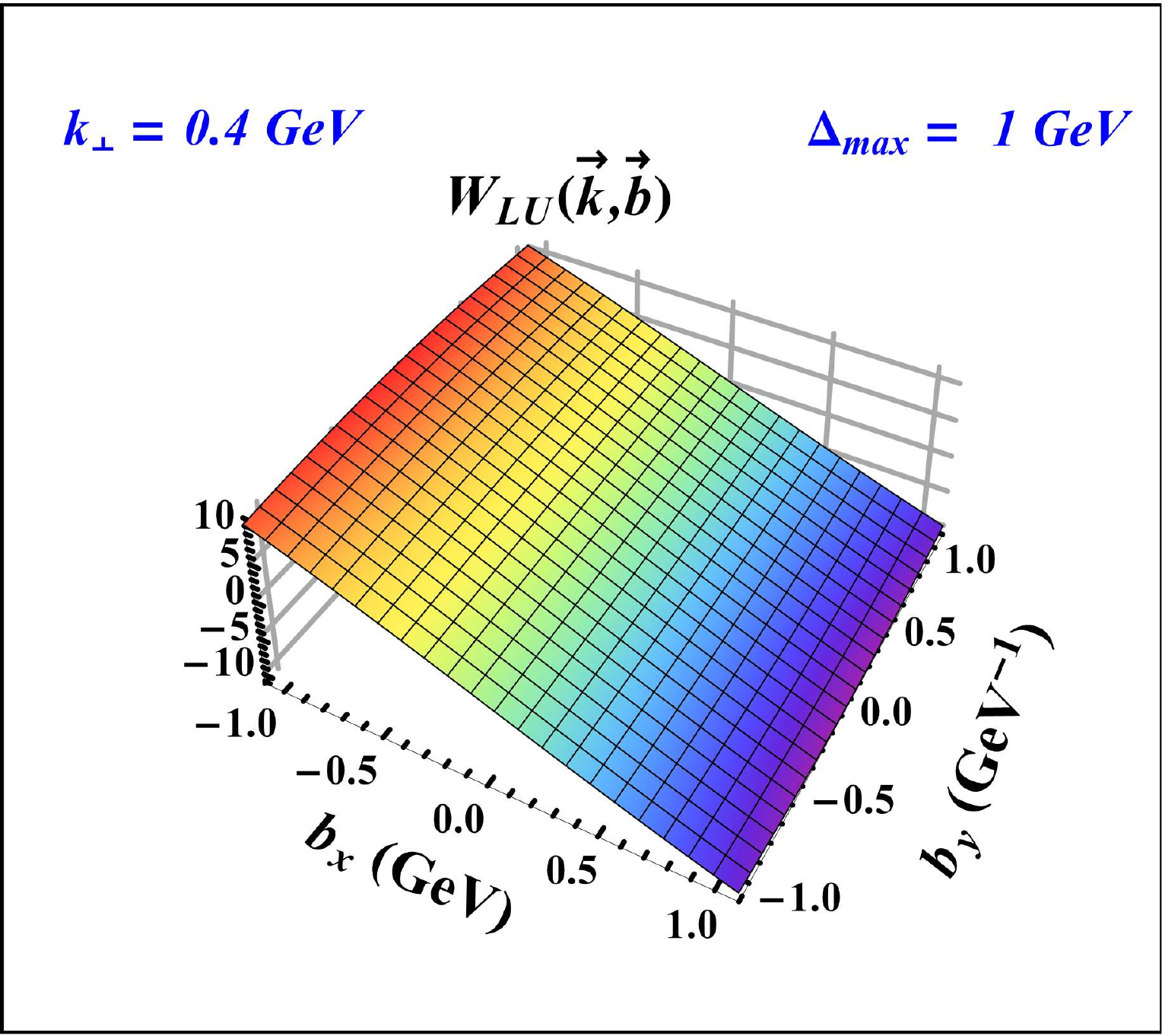}
\hspace{0.1cm}
\tiny{(b)}\includegraphics[width=7.8cm,height=6cm,clip]{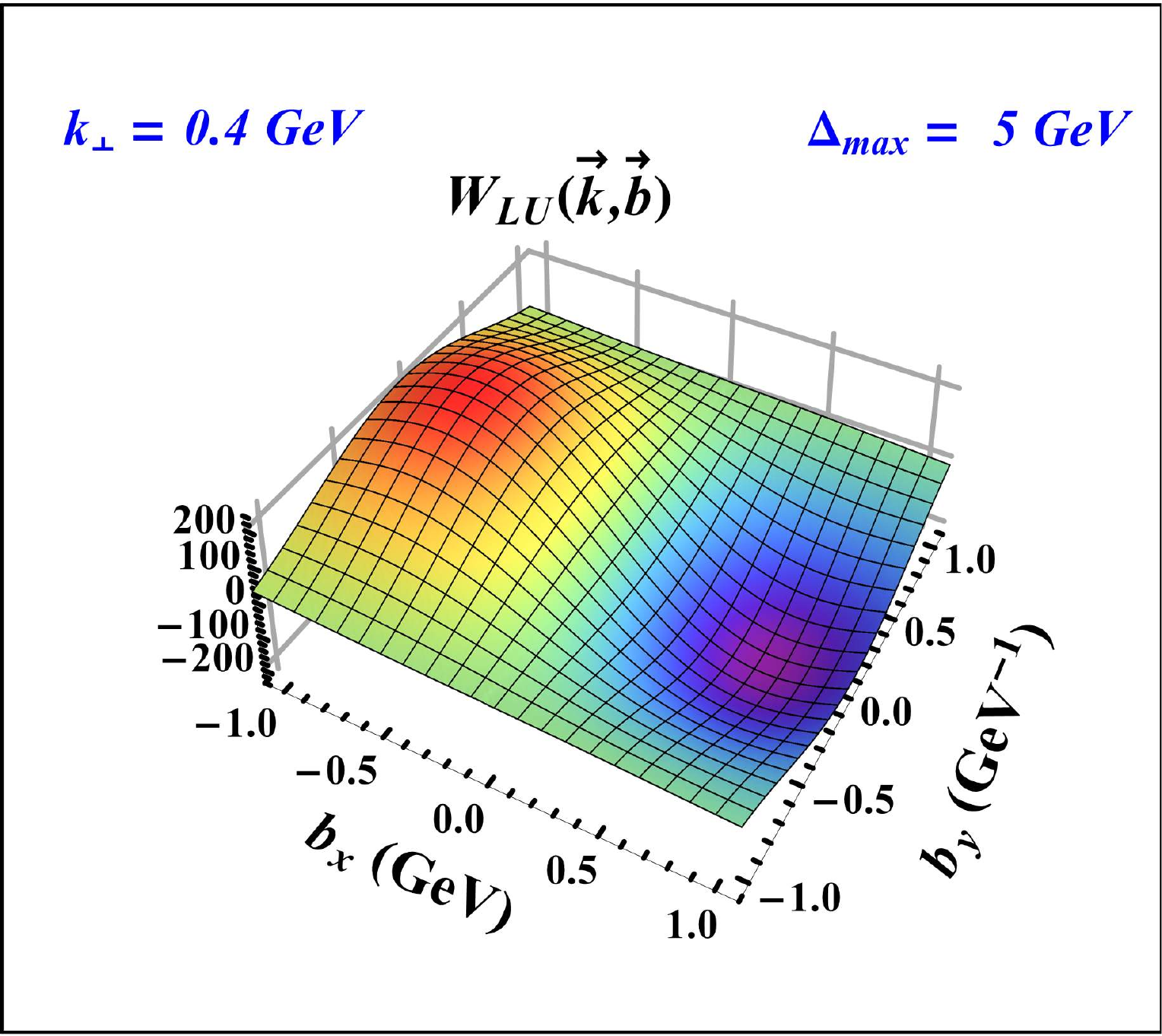}
\end{minipage}
\begin{minipage}[c]{1\textwidth}
\tiny{(c)}\includegraphics[width=7.8cm,height=6cm,clip]{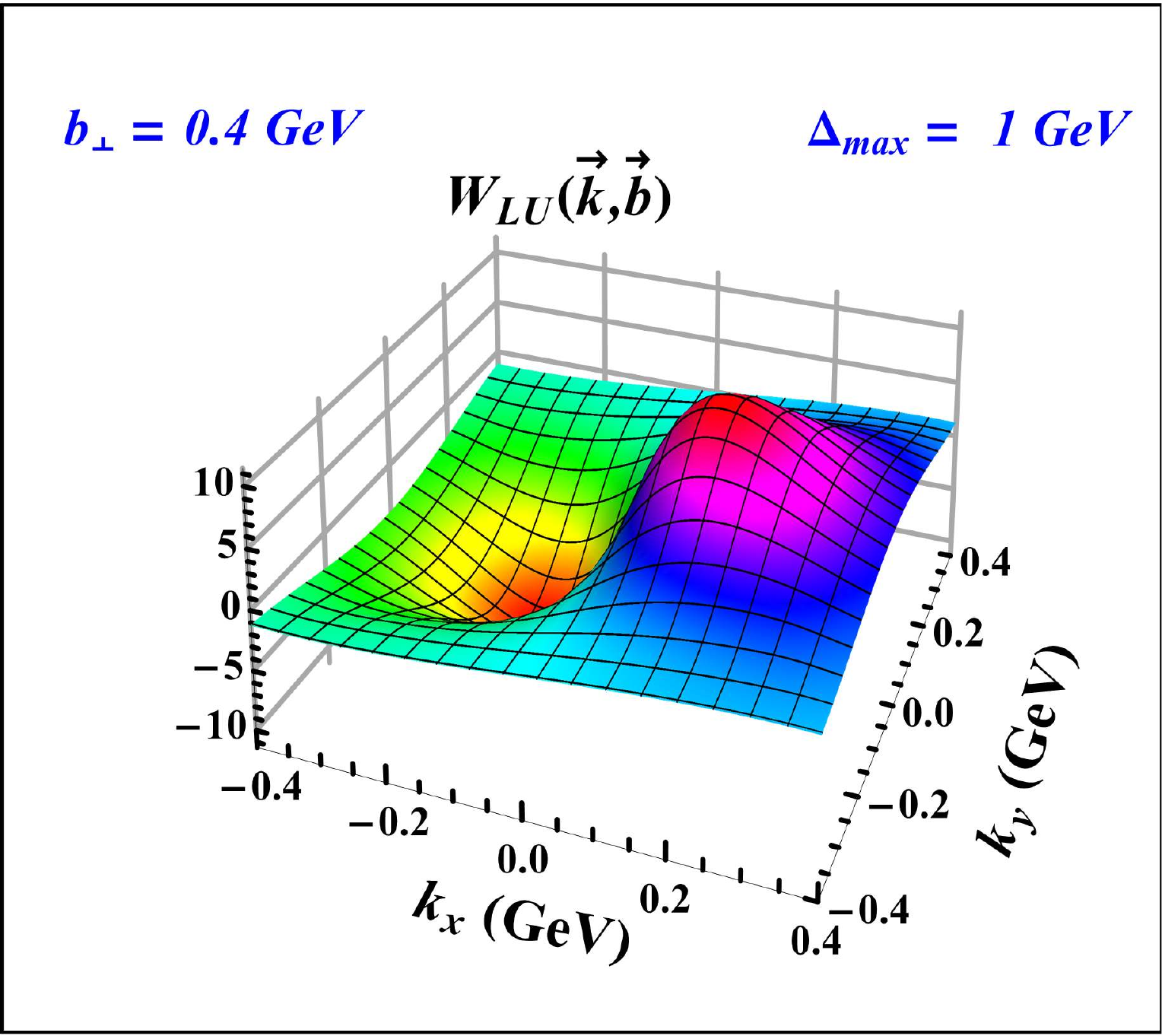}
\hspace{0.1cm}
\tiny{(d)}\includegraphics[width=7.8cm,height=6cm,clip]{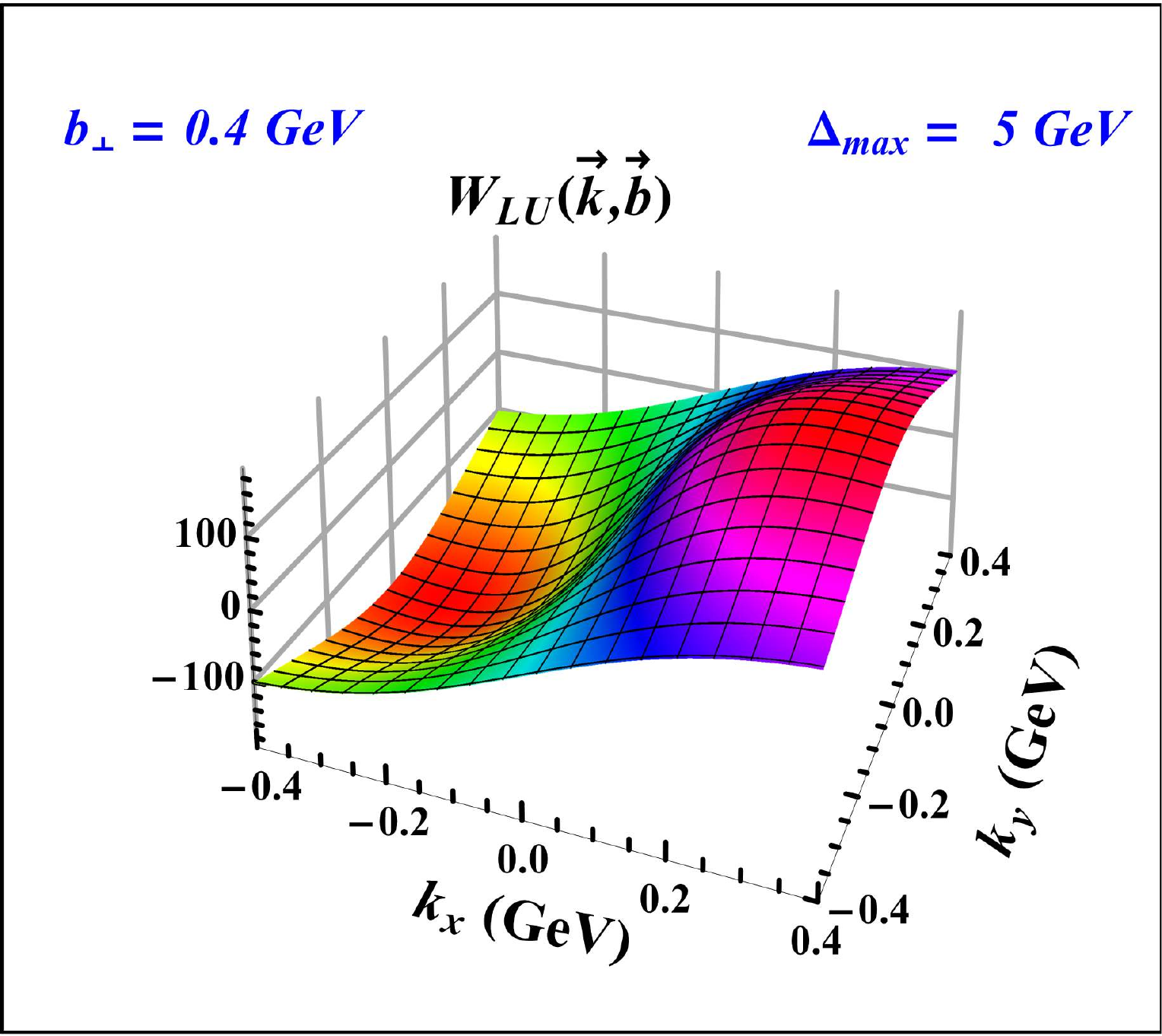}
\end{minipage}
\begin{minipage}[c]{1\textwidth}
\tiny{(e)}\includegraphics[width=7.8cm,height=6cm,clip]{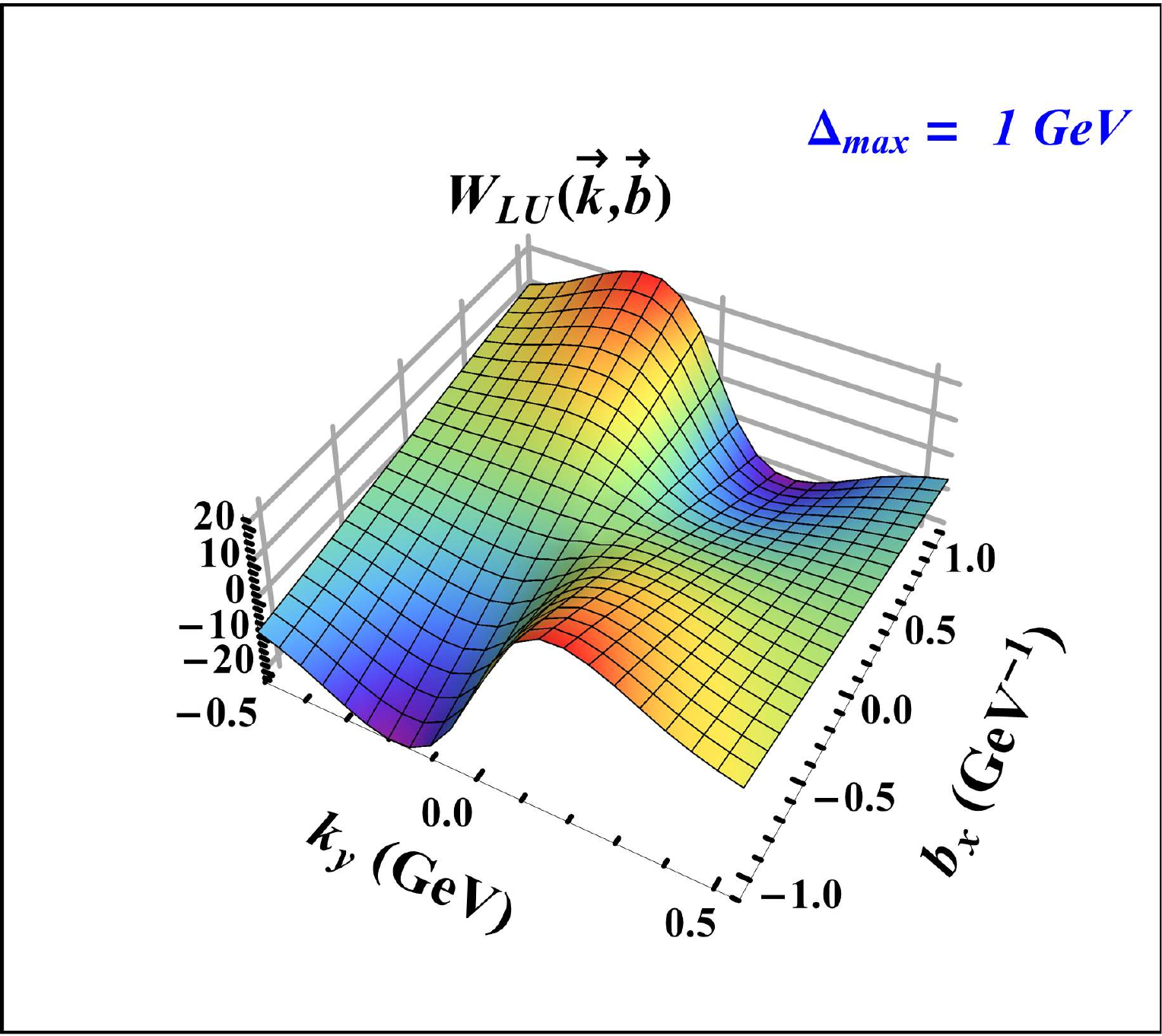}
\hspace{0.1cm}
\tiny{(f)}\includegraphics[width=7.8cm,height=6cm,clip]{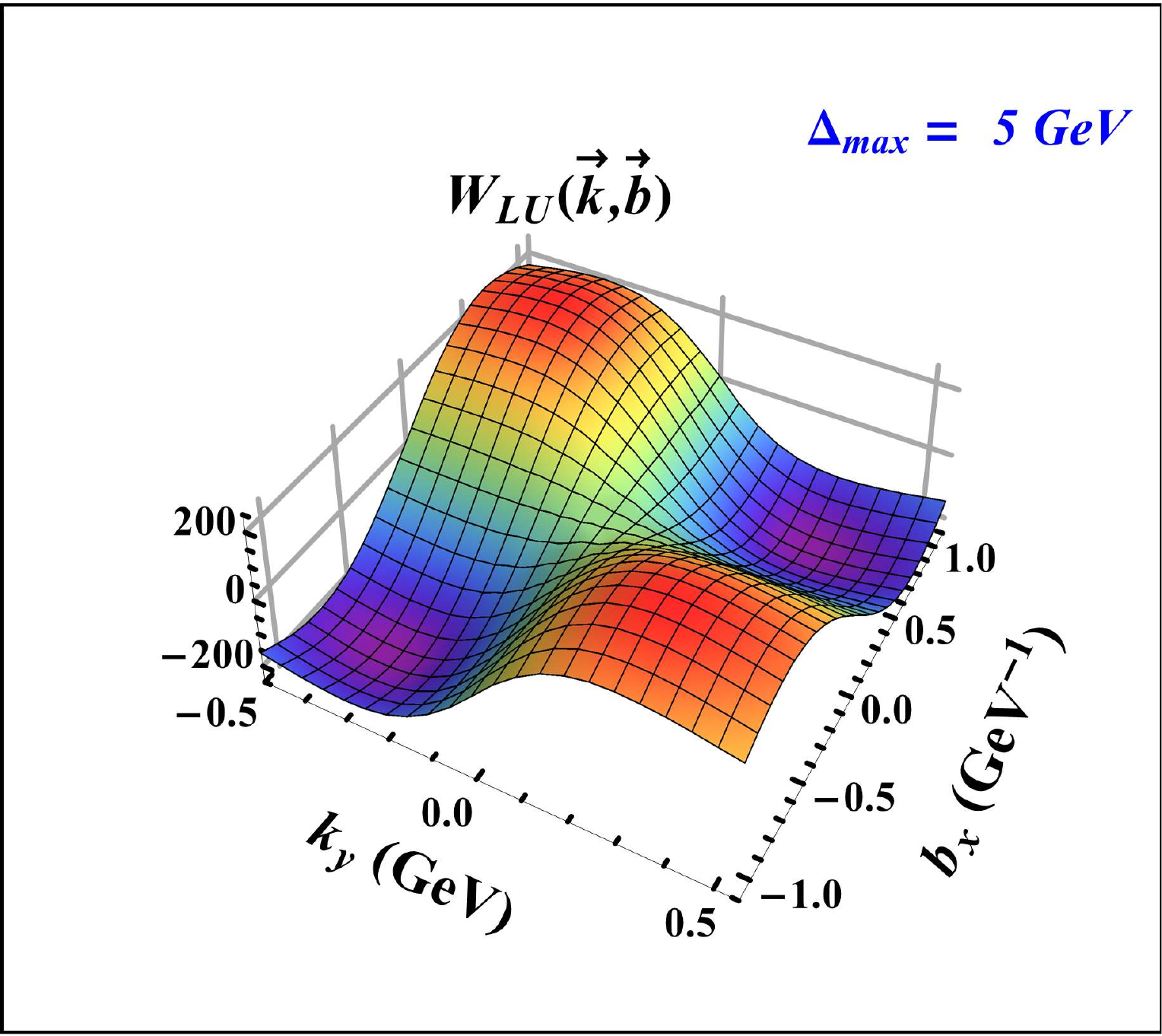}
\end{minipage}
\caption{\label{fig2}(Color online)
3D plots of the Wigner distributions $W^{LU}$. Plots (a) and (b) are in $b$ space with $k_\perp = 0.4$ GeV.
Plots (c) and (d) are in $k$ space with $b_\perp = 0.4$ $ \mathrm{GeV}^{-1}$.
Plots (e) and (f) are in mixed space where $k_x$ and $b_y$ are integrated.
All the plots on the left panel (a,c,e) are for $\Delta_{max} = 1.0$ GeV. Plots on the right panel (b,d,f) are for $\Delta_{max} = 5.0$ GeV.
For all the plots we kept $m = 0.33$ GeV, integrated out the $x$ variable and we took $\vec{k_\perp} = k \hat{j}$ 
and $\vec{b_\perp} = b \hat{j}$.  }
\end{figure}

\begin{figure}[!htp]
\begin{minipage}[c]{1\textwidth}
\tiny{(a)}\includegraphics[width=7.8cm,height=6cm,clip]{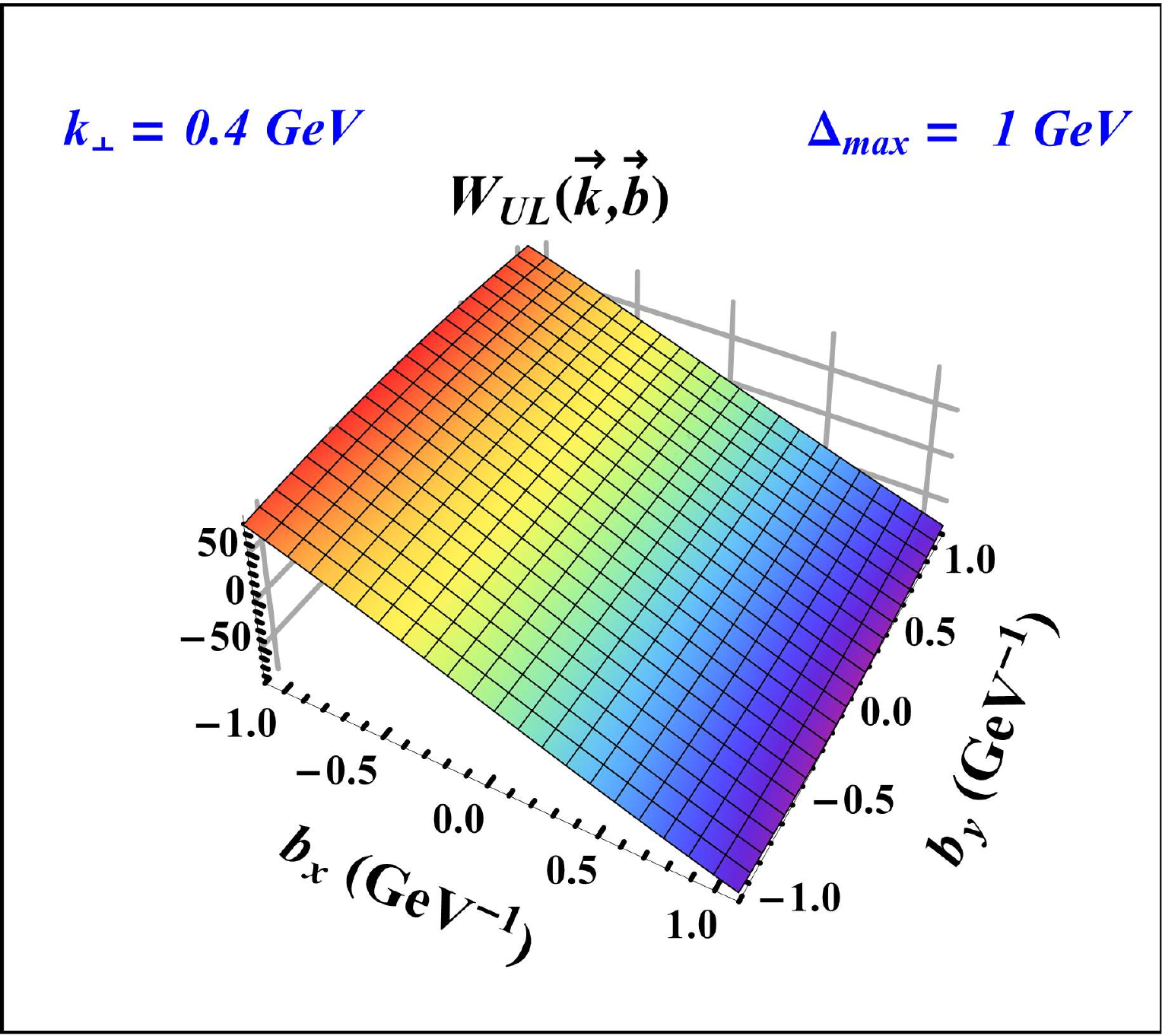}
\hspace{0.1cm}
\tiny{(b)}\includegraphics[width=7.8cm,height=6cm,clip]{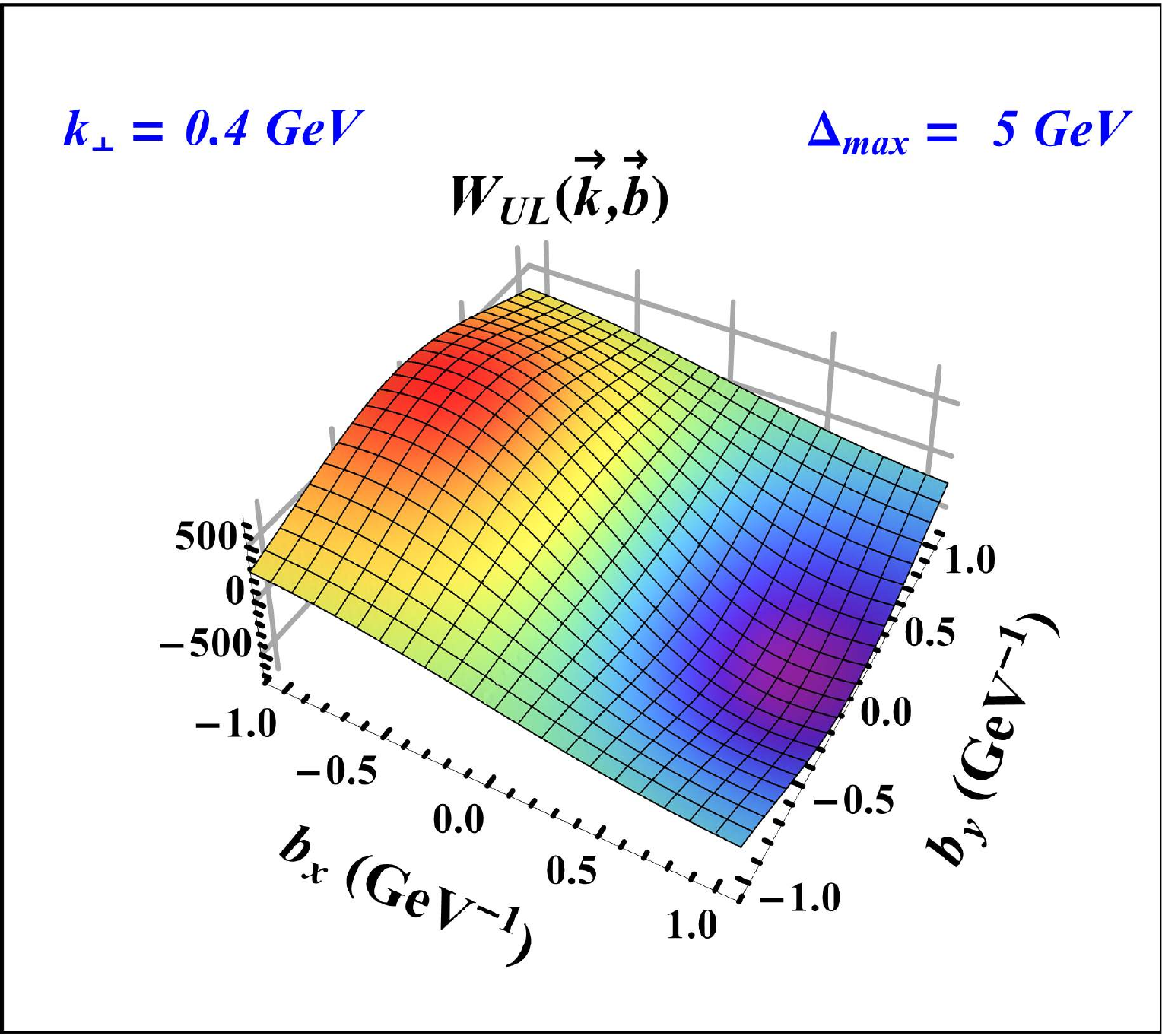}
\end{minipage}
\begin{minipage}[c]{1\textwidth}
\tiny{(c)}\includegraphics[width=7.8cm,height=6cm,clip]{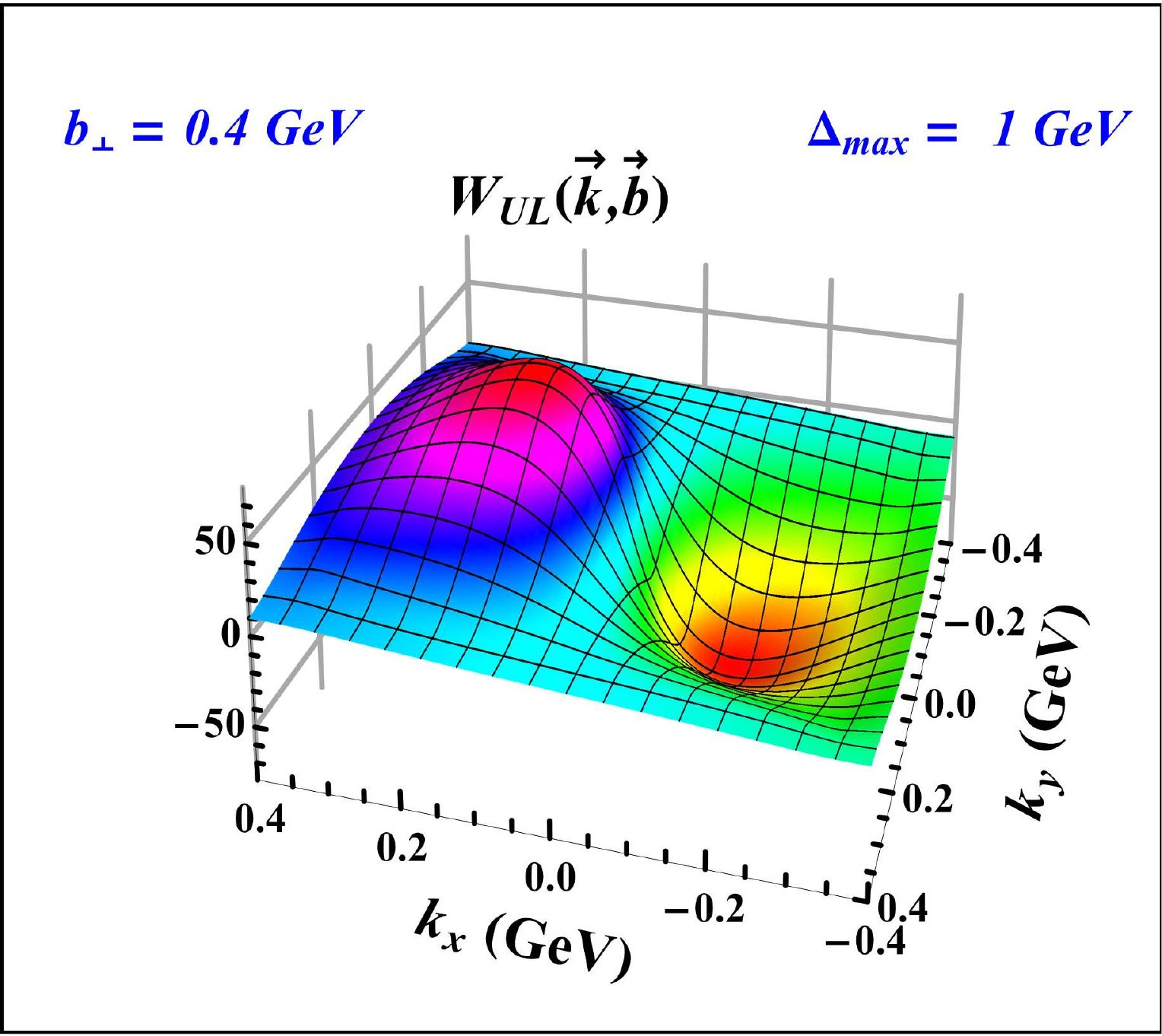}
\hspace{0.1cm}
\tiny{(d)}\includegraphics[width=7.8cm,height=6cm,clip]{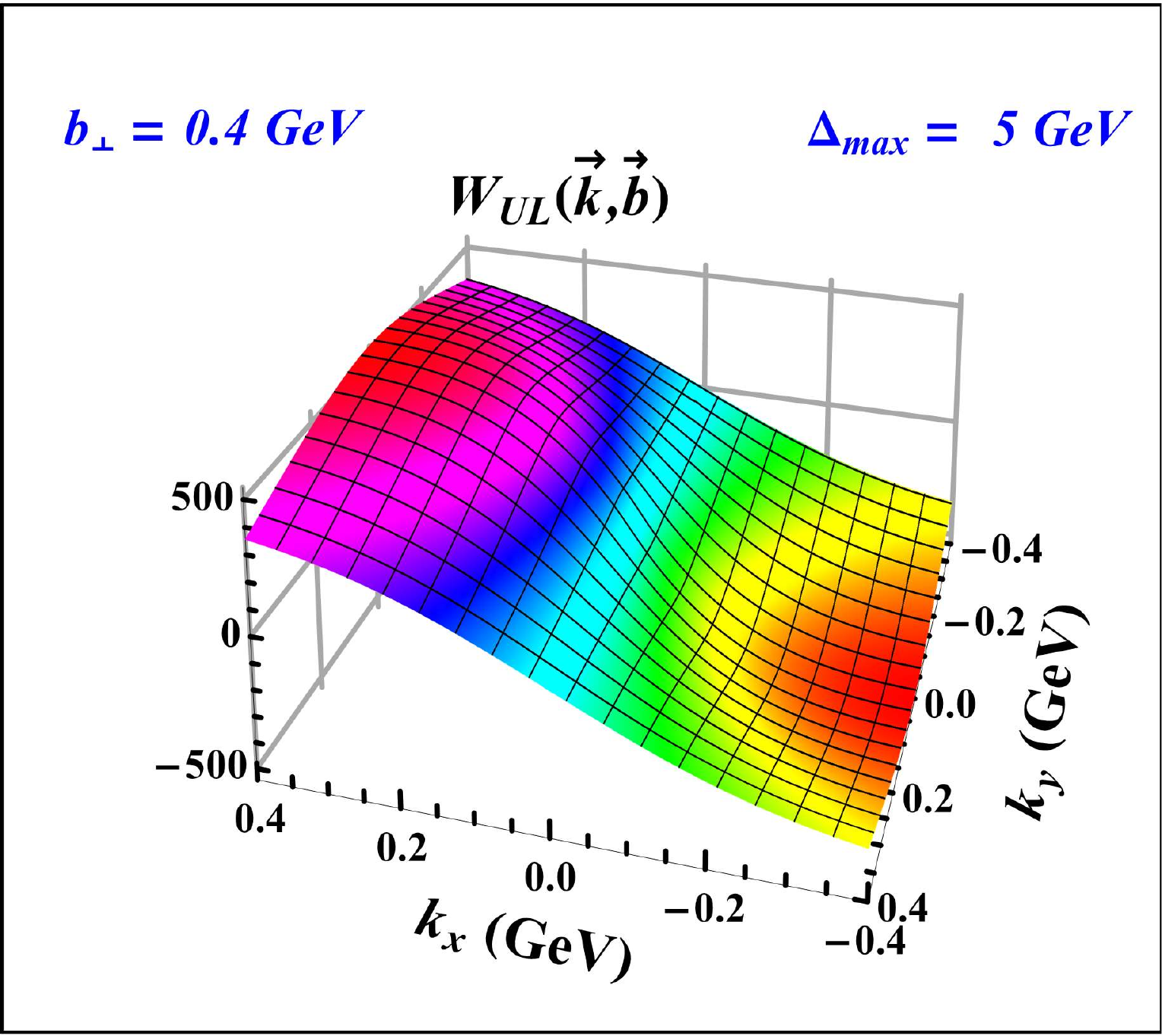}
\end{minipage}
\begin{minipage}[c]{1\textwidth}
\tiny{(e)}\includegraphics[width=7.8cm,height=6cm,clip]{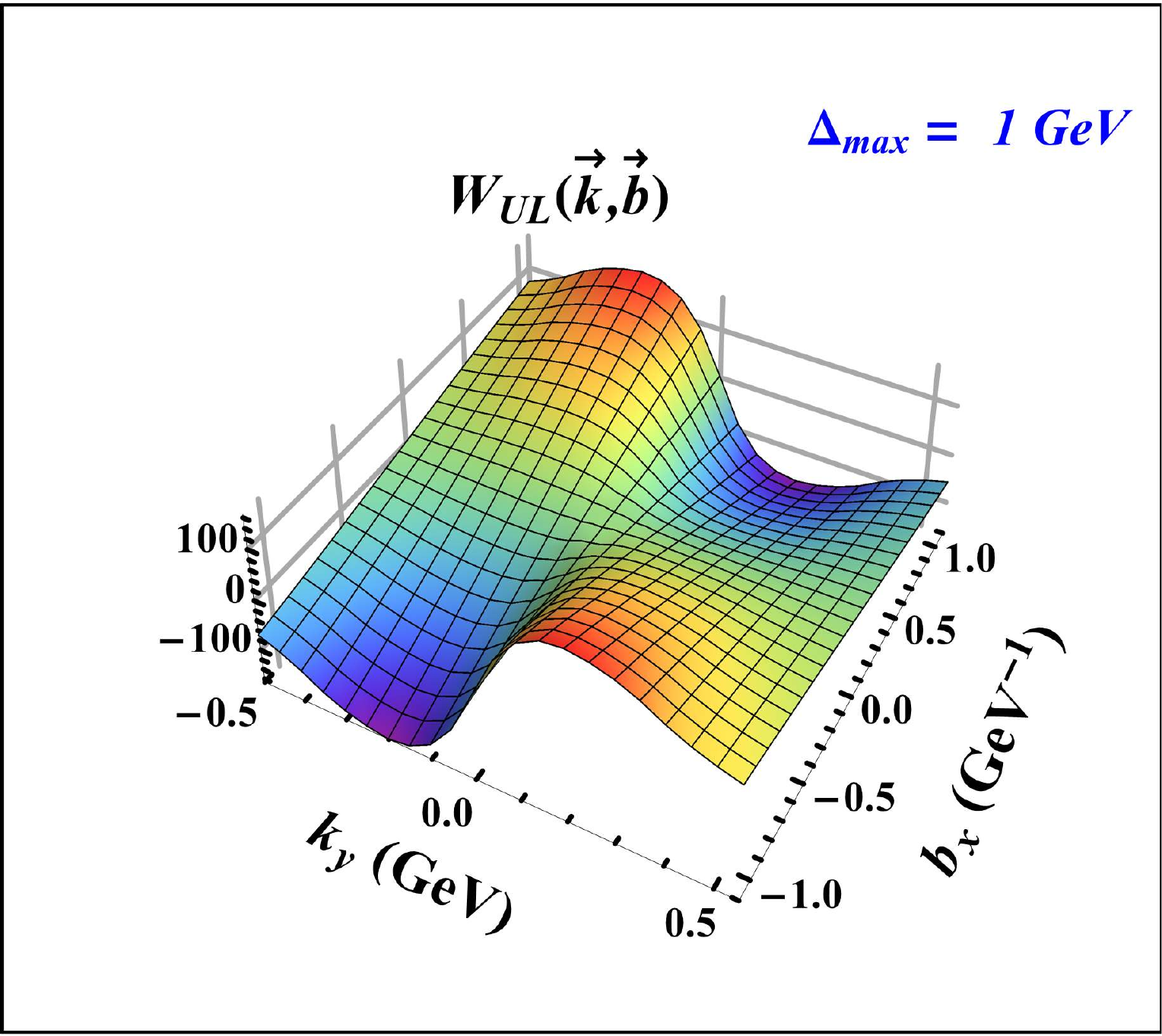}
\hspace{0.1cm}
\tiny{(f)}\includegraphics[width=7.8cm,height=6cm,clip]{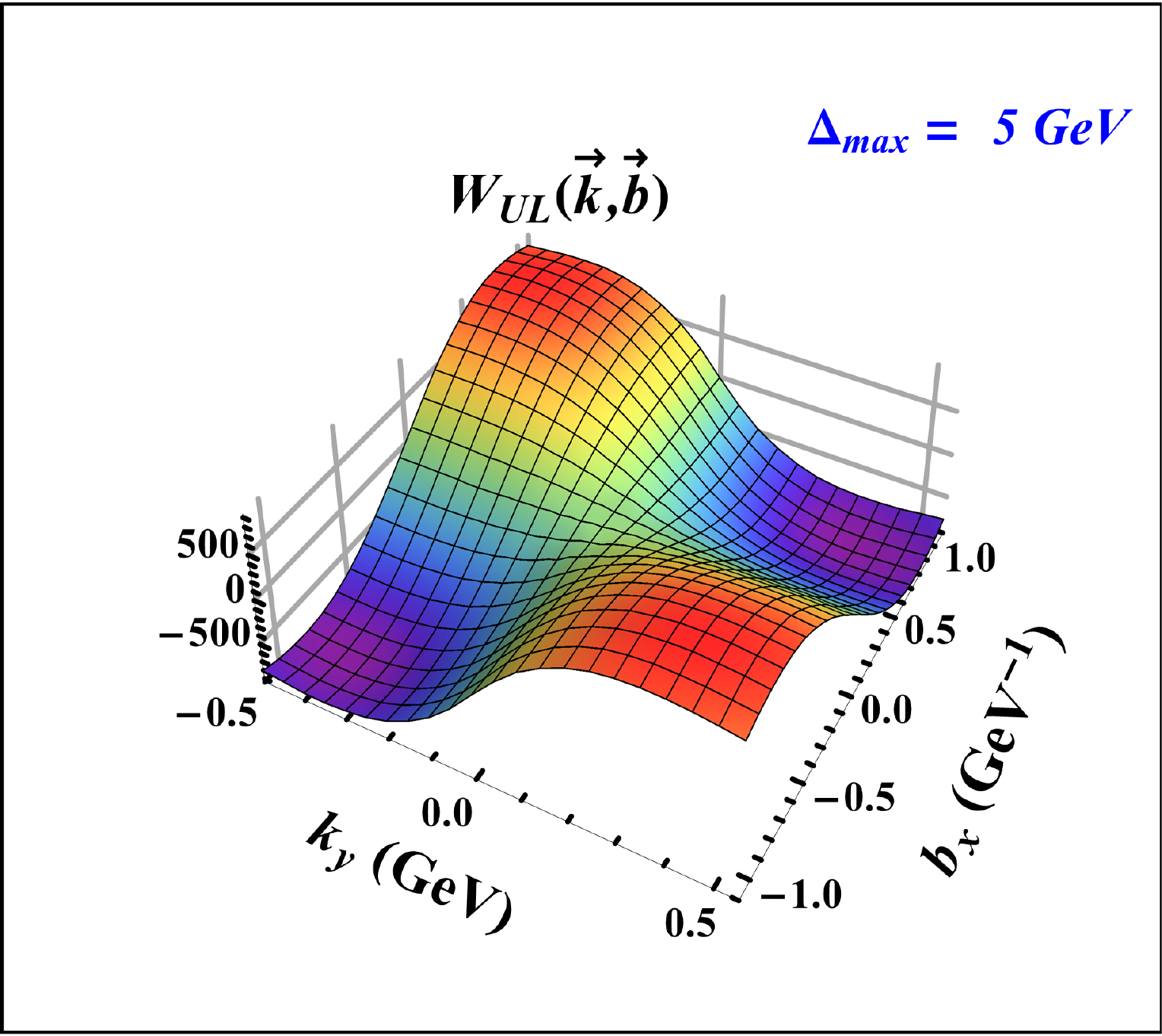}
\end{minipage}
\caption{\label{fig3}(Color online)
3D plots of the Wigner distributions $W^{UL}$. Plots (a) and (b) are in $b$ space with $k_\perp = 0.4$ GeV.
Plots (c) and (d) are in $k$ space with $b_\perp = 0.4$ $ \mathrm{GeV}^{-1}$.
Plots (e) and (f) are in mixed space where $k_x$ and $b_y$ are integrated.
All the plots on the left panel (a,c,e) are for $\Delta_{max} = 1.0$ GeV. Plots on the right panel (b,d,f) are for $\Delta_{max} = 5.0$ GeV.
For all the plots we kept $m = 0.33$ GeV, integrated out the $x$ variable and we took $\vec{k_\perp} = k \hat{j}$ 
and $\vec{b_\perp} = b \hat{j}$.  }
\end{figure}

\begin{figure}[!htp]
\begin{minipage}[c]{1\textwidth}
\tiny{(a)}\includegraphics[width=7.8cm,height=6cm,clip]{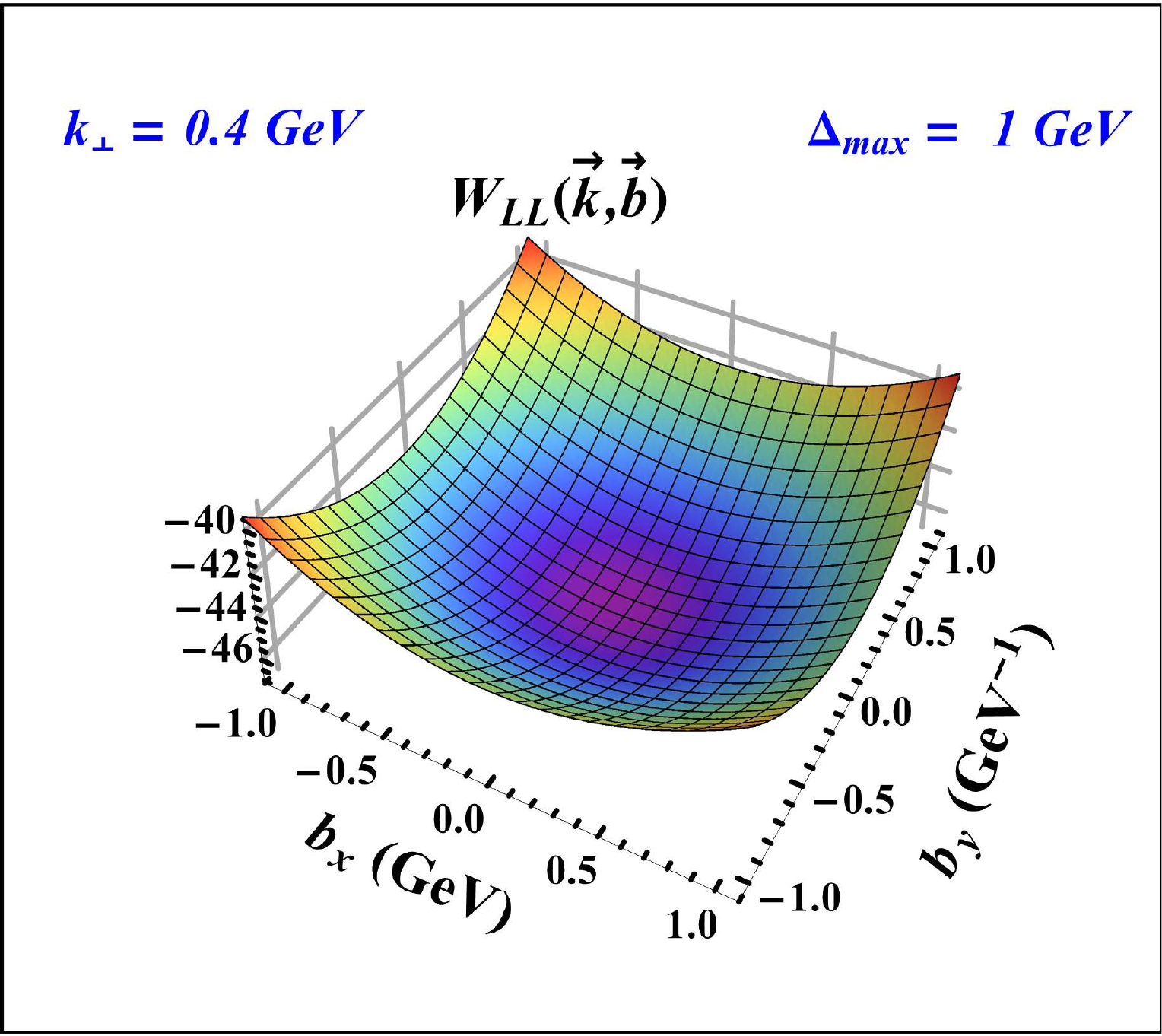}
\hspace{0.1cm}
\tiny{(b)}\includegraphics[width=7.8cm,height=6cm,clip]{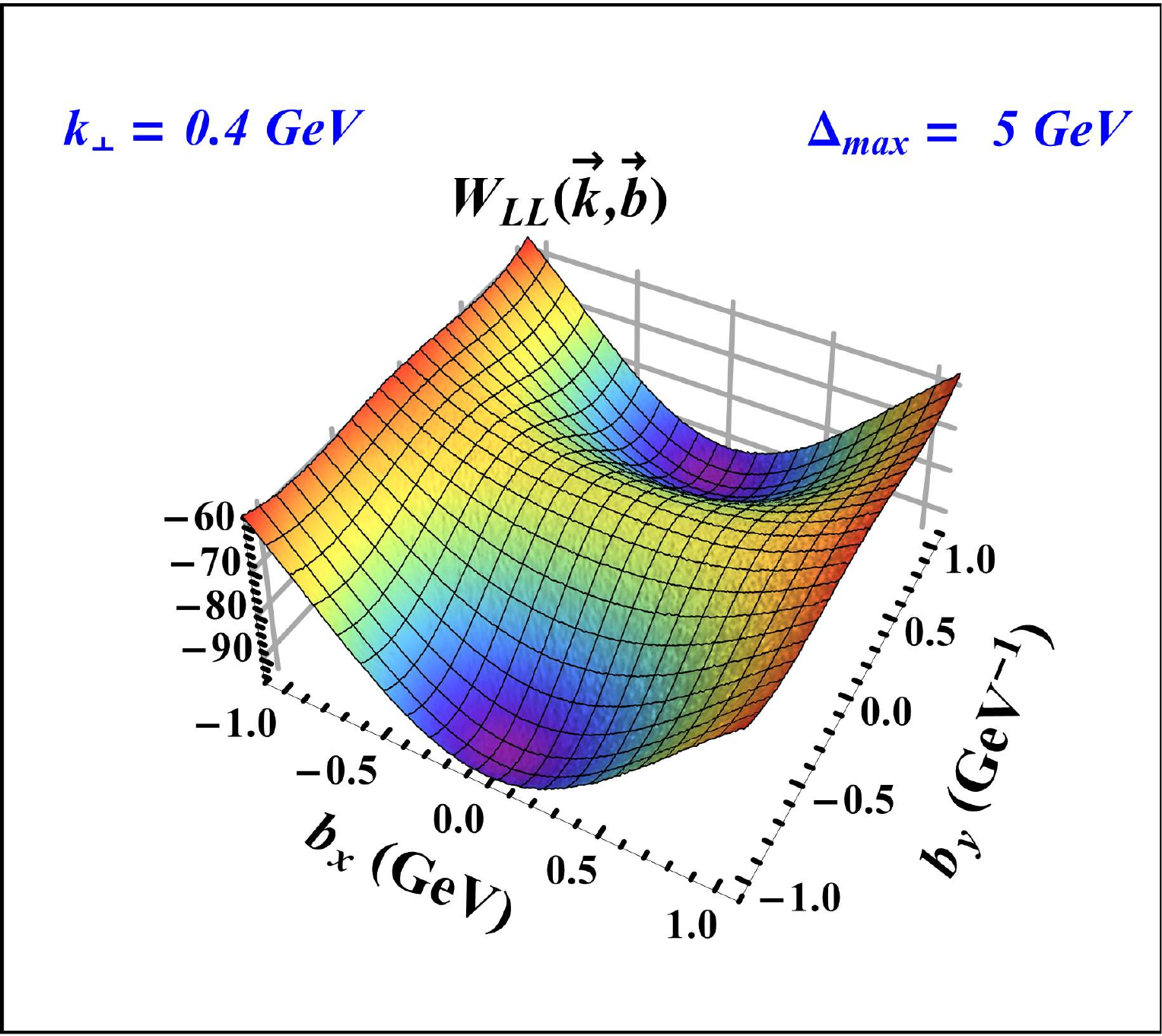}
\end{minipage}
\begin{minipage}[c]{1\textwidth}
\tiny{(c)}\includegraphics[width=7.8cm,height=6cm,clip]{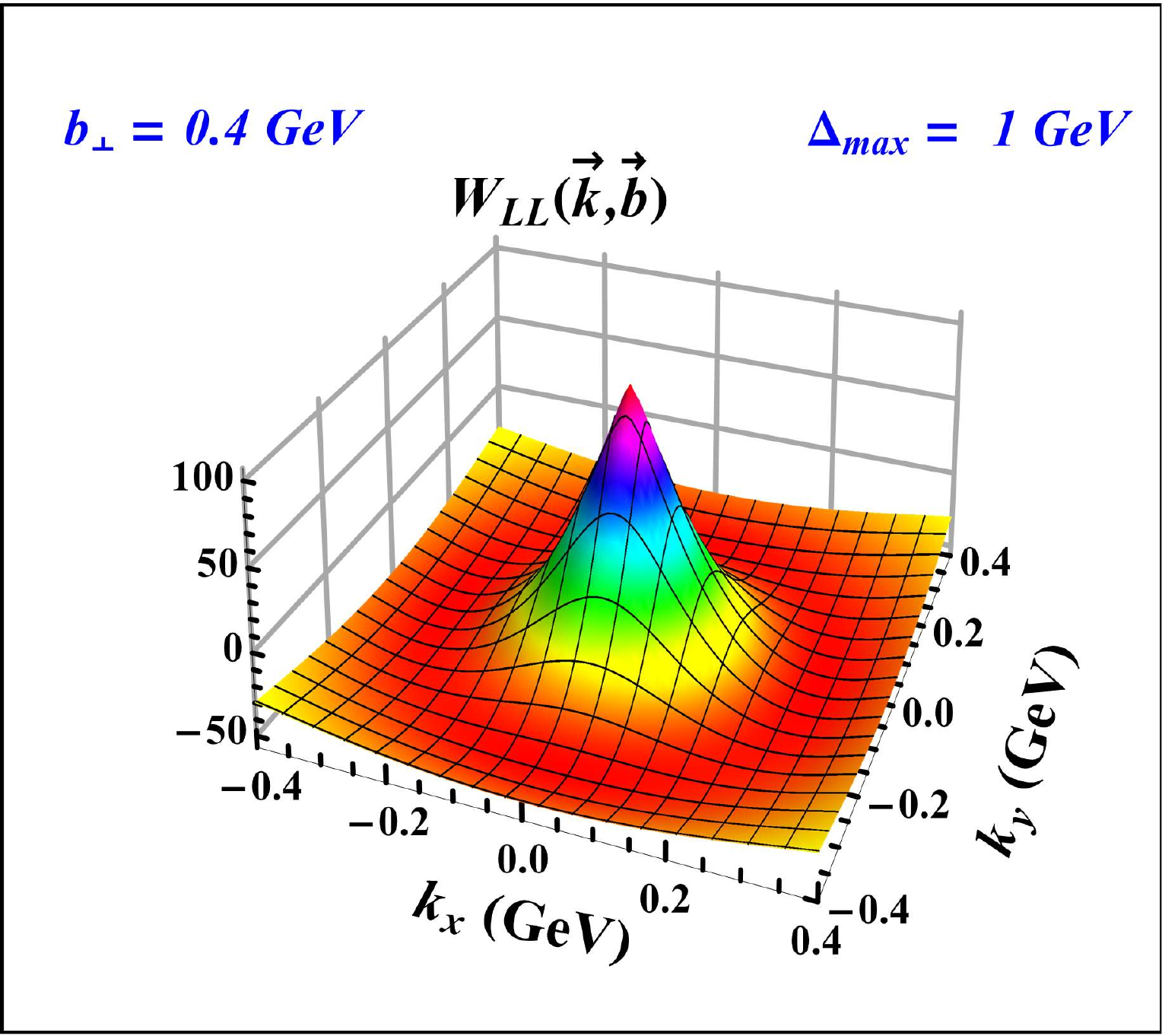}
\hspace{0.1cm}
\tiny{(d)}\includegraphics[width=7.8cm,height=6cm,clip]{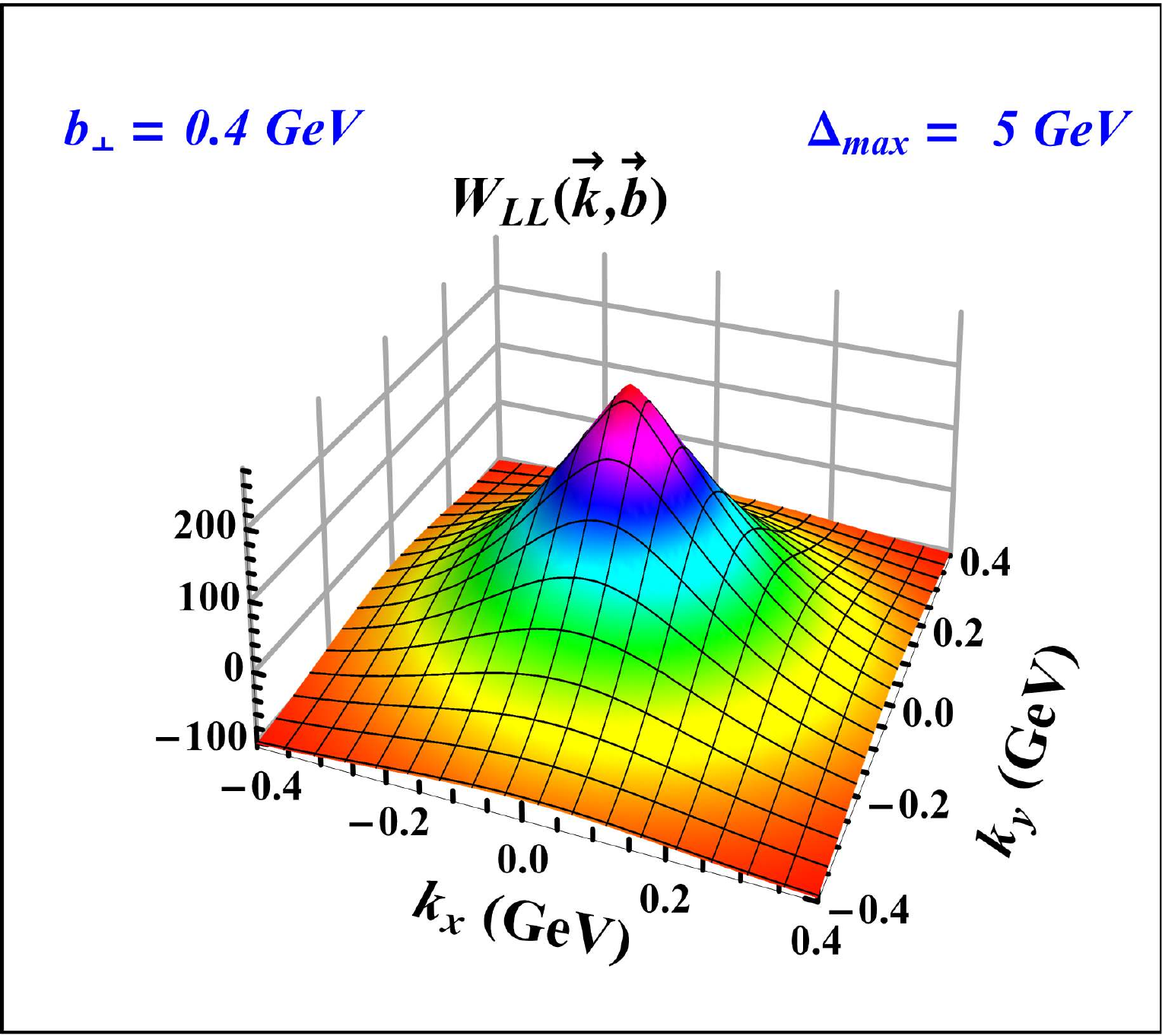}
\end{minipage}
\begin{minipage}[c]{1\textwidth}
\tiny{(e)}\includegraphics[width=7.8cm,height=6cm,clip]{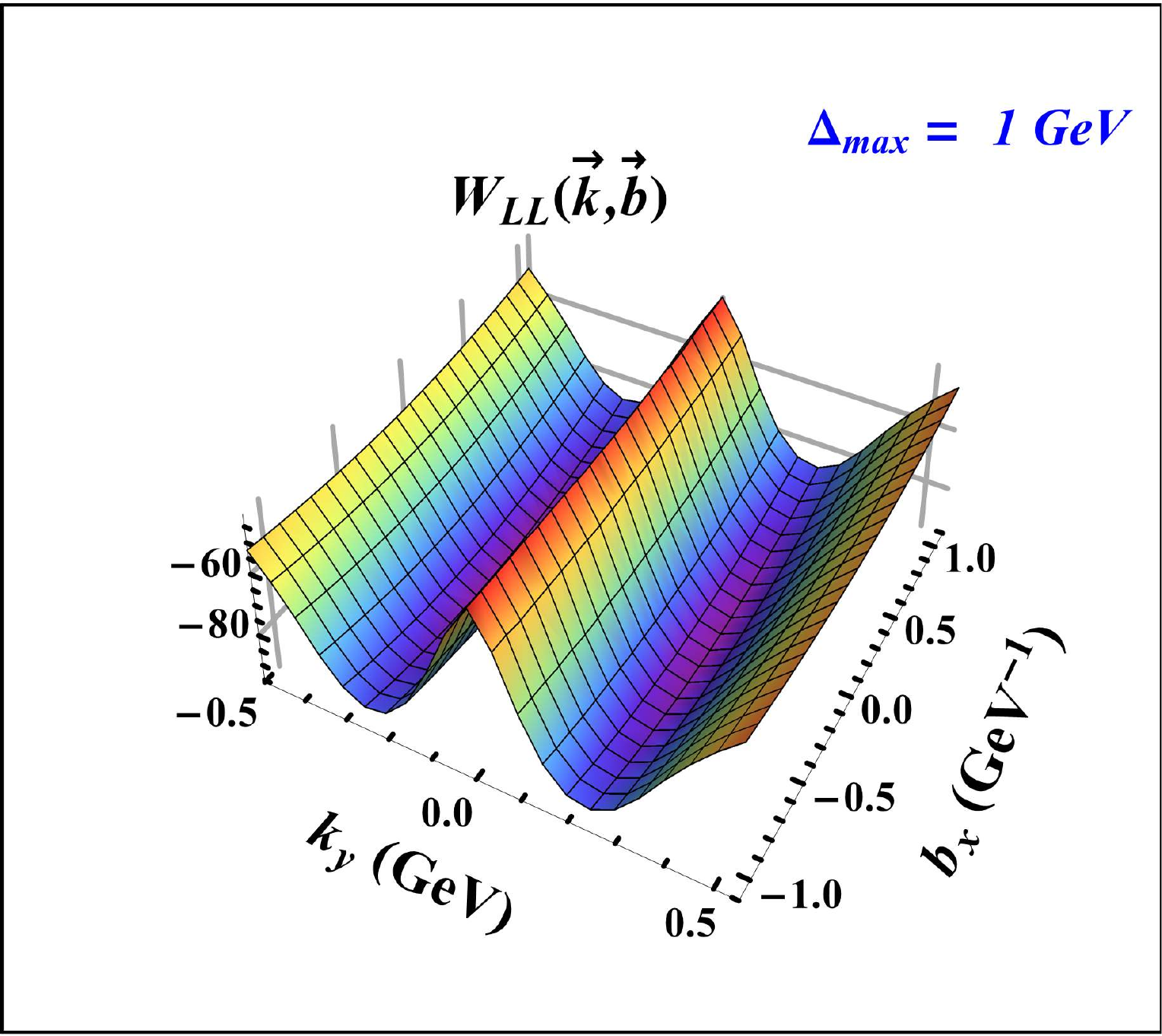}
\hspace{0.1cm}
\tiny{(f)}\includegraphics[width=7.8cm,height=6cm,clip]{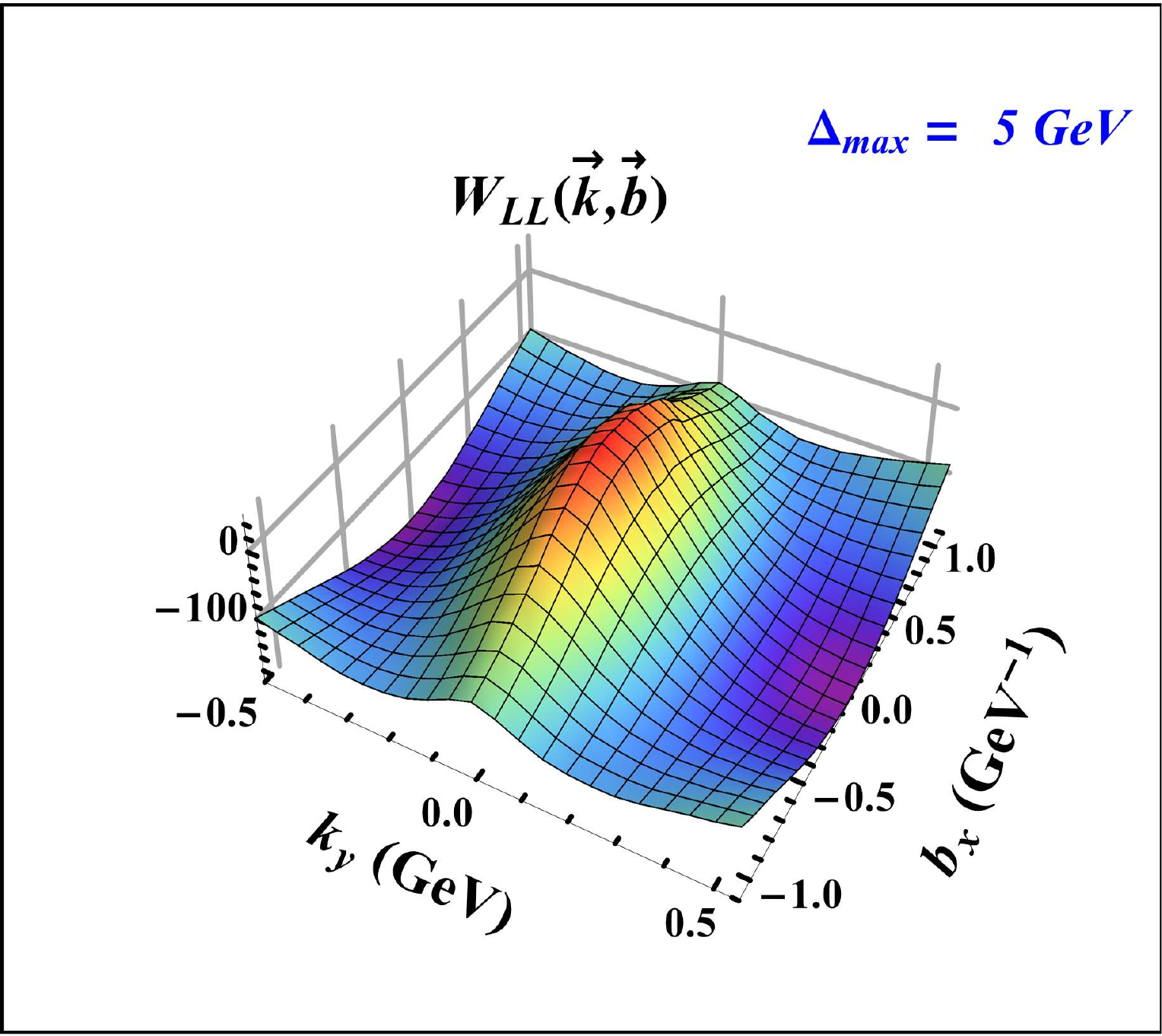}
\end{minipage}
\caption{\label{fig4}(Color online)
3D plots of the Wigner distributions $W^{LL}$. Plots (a) and (b) are in $b$ space with $k_\perp = 0.4$ GeV.
Plots (c) and (d) are in $k$ space with $b_\perp = 0.4$ $ \mathrm{GeV}^{-1}$.
Plots (e) and (f) are in mixed space where $k_x$ and $b_y$ are integrated.
All the plots on the left panel (a,c,e) are for $\Delta_{max} = 1.0$ GeV. Plots on the right panel (b,d,f) are for $\Delta_{max} = 5.0$ GeV.
For all the plots we kept $m = 0.33$ GeV, integrated out the $x$ variable and we took $\vec{k_\perp} = k \hat{j}$ 
and $\vec{b_\perp} = b \hat{j}$.  }
\end{figure}

 \section{Gluon GTMDs and Orbital Angular Momentum}
 
 \noindent
In order to calculate the gluon GTMDs we use the parametrization as shown in \cite{gluon_gtmd} 
where the authors have shown that the correlators like in Eq.~(\ref{eq1}) can in general be written as

\be 
W^{\it{O}}_{\Lambda',\Lambda} = \int \frac{dz^{-}d^{2}z_{\perp}}{2(2\pi)^3} 
e^{i\frac{xP^{+}}{2}z^- - i\vec{k}_{\perp}.\vec{z}_{\perp}} \langle p',\Lambda' \mid O(z)\mid p,\Lambda \rangle = 
\bar{u}(p',\Lambda' ) M^{O}u(p,\Lambda );
\label{amp}\ee

\noindent
where $O(z)$ stands for the relevant quark or gluon operators and $M^{O}$ stands for the matrix in Dirac space 
with $O$ determined by the corresponding quark or gluon operator. The amplitude shown in Eq.~(\ref{amp}) takes the 
following generic structure when the momentum transfer is purely in the
transverse direction :

\be 
W^{\Delta S_z,c_P}_{\Lambda',\Lambda} = \frac{\bar{u}(p',\Lambda')M^{\Delta S_z,c_P}u(p,\Lambda)}{2P^+};
\label{eq3}\ee \\

\noindent
where  $c_P$ is the parity coefficient of the partonic operator and  $\Delta S_z$ is the spin flip number given by 
$\Delta S_z = \lambda' - \lambda + \Delta L_z$ such that $\lambda (\lambda')$ is the initial (final) parton light front helicity 
and $\Delta L_z$ is the eigenvalue of the operator $\Delta \hat{L}_z= \hat{L}_z - \hat{L'}_z$. Also for twist-2 partonic 
operators we get $\Delta L_z = 0$ and hence $\Delta S_z = \lambda' - \lambda$.\\

\noindent
The gluon operators appearing in Eq.~(\ref{eq1}) corresponds to the case when $\Delta S_z=0$ and  $c_P = \pm 1$ as shown in eq (3.12), eq (3.42) and eq (3.43) of Eq.~(\ref{gtmd}). 
So the relevant parameterization of the gluon GTMDs which correspond to $\Gamma^{ij} = \delta^{ij}$ in
Eq.~(\ref{eq1}) is:

\be
M^{0,+} = \Big(\frac{M}{P^+}\Big)^{t-1} \Big[  \gamma^+ \Big( S^{0,+}_{t,ia}+ \gamma_5 \frac{i \epsilon_{T}^{k_{T}\Delta_{T}}}{M^2}S^{0,+}_{t,ib} \Big)  + i\sigma^{j+}\Big( \frac{k_{T}^j}{M}
P^{0,+}_{t,ia} +\frac{\Delta_{T}^j}{M}
P^{0,+}_{t,ib} \Big)\Big];
\label{gtmd}
\ee 
\noindent
where
\be
\epsilon_{T}^{ab}&=&\epsilon_{T}^{ij}a^ib^j; \nn \\
\epsilon_{T}^{ij}&=&\epsilon^{-+12}=+1.\nn \\
\nn \ee

\noindent
t+1 is defined as the twist of the operator in \cite{gluon_gtmd}, so for twist 2 we take $t=1$.\\
Comparing and solving Eq.~(\ref{eq1}) and Eq.~(\ref{eq3}), we get the following expression for the gluon
GTMDs:

\be
S^{0,+}_{1,ia} &=& \frac{-2N}{D(q_{\perp})D(q'_{\perp})}\Bigg[  \frac{m^2x^4 + k_{\perp}^2 (x^2-2x+2)}{x^3(x-1)^2} - \frac{\Delta_{\perp}^2 (x^2-2x+2)}{4x^3}  \Bigg]; \nn \\ \nn \\
S^{0,+}_{1,ib} &=&   \frac{2N}{D(q_{\perp})D(q'_{\perp})}\Bigg[      \frac{m^2(2-x)}{(1-x)x^2}      \Bigg]; \nn \\ \nn \\
P^{0,+}_{1,ia} &=& \frac{2N}{D(q_{\perp})D(q'_{\perp})}\Bigg[  \frac{m^2\Delta_{\perp}^2}{x(k_2\Delta_1-k_1\Delta_2)}   \Bigg]; \nn \\ \nn \\
P^{0,+}_{1,ib} &=& \frac{2N}{D(q_{\perp})D(q'_{\perp})}\Bigg[  \frac{-m^2(k_\perp.\Delta_\perp)}{x(k_2\Delta_1-k_1\Delta_2)}   \Bigg].
\ee \\

The relations of these GTMDs with that in \cite{metz} are:-
 
\be
S^{0,+}_{1,ia} &=& F_{1,1}^g;\nn \\
S^{0,+}_{1,ib} &=& F_{1,4}^g;\nn \\
P^{0,+}_{1,ia} &=& F_{1,2}^g;\nn \\
P^{0,+}_{1,ib} &=& -\frac{F_{1,1}^g}{2}+F_{1,3}^g.\nn \\
\label{rel}
\ee

So for the gluon case we get $F_{1,4}^g$ as shown below and this relation agrees with that in \cite{other}.

\be
F_{1,4}^g = \frac{N}{D(q_{\perp})D(q'_{\perp})}\Bigg[ \frac{m^2(2-x)}{(1-x)x^2}      \Bigg].
\ee \\
 
\noindent
From this, we can calculate the gluon canonical OAM since the canonical OAM is related to the 
GTMD $F_{1,4}$ as follows, similar to quarks \cite{lorce,OAM3,hatta1}
\be
l^{g}_{z} = -\int dx d^{2}k_{\perp} \frac{k_{\perp}^2}{m^2} F_{1,4}^g.
\ee

This gives,  

\be \label{sl}
l^{g}_{z} = -  N \int dx (1-x)(2-x)\Big[ I_{1} - m^2x^2 I_{2}\Big ];
 \ee%

where,
 
\begin{align}
  I_{1} &=   \int 
 \frac{d^{2} k_{\perp}}{m^2x^2 +(k_{\perp})^2}  = \pi log\Bigg[\frac{Q^2+m^2x^2}{\mu^2+m^2
x^2}\Bigg];\nn \\
  I_{2} &=  \int  \frac{d^{2} k_{\perp}}{\Big(m^2x^2 +(k_{\perp})^2\Big)^2} =
  \frac{\pi}{(m^2 x^2)};\nn\\ 
 \nn 
 \end{align}
 Here $Q$ and $\mu$ are the upper and lower limits of the $k_\perp$ integration
respectively.\\

\noindent
For the case when $\Delta S_z=0$ and  $c_P = -1$ which corresponds to 
$\Gamma^{ij} = -i\epsilon^{ij}_{\perp}$ in Eq.~(\ref{eq1}) the relevant gluon parameterization is given by \cite{gluon_gtmd}: 

\be
M^{0,-} = \Big(\frac{M}{P^+}\Big)^{t-1} \Big[  \gamma^+ \gamma_5  \Big( S^{0,-}_{t,ia}+ 
\gamma_5 \frac{i \epsilon_{T}^{k_{T}\Delta_{T}}}{M^2}S^{0,-}_{t,ib} \Big)  + i\sigma^{j+}\gamma_5\Big( \frac{k_{T}^j}{M}
P^{0,-}_{t,ia} +\frac{\Delta_{T}^j}{M}
P^{0,-}_{t,ib} \Big)\Big].
\label{gtmd1}
\ee \\
\noindent

Again by solving Eq.~(\ref{gtmd1}) and Eq.~(\ref{eq1}) we get the corresponding GTMDs at twist two.

\be
S^{0,-}_{1,ia} &=& \frac{N}{D(q_{\perp})D(q'_{\perp})}\Bigg[  \frac{4 k_{\perp}^2 (x-2)}{2(x-1)^2x^2} + \frac{\Delta_{\perp}^2}{(x-1)^2x^2} - \frac{4m^2x^2 +\Delta_{\perp}^2 (x^2-4x+5) }{2(x-1)^2 x} \Bigg]; \nn \\ \nn \\
S^{0,-}_{1,ib} &=&  \frac{N}{D(q_{\perp})D(q'_{\perp})}\Bigg[    \frac{2(x^2-2x+2)m^2}{x^3(1-x)}   \Bigg]; \nn \\ \nn \\
P^{0,-}_{1,ia} &=& \frac{N}{D(q_{\perp})D(q'_{\perp})}\Bigg[ \frac{4(k_{\perp}.\Delta_\perp)m^2}{x(x-1)(k_2\Delta_1-k_1\Delta_2)} \Bigg]; \nn \\ \nn \\
P^{0,-}_{1,ib} &=& \frac{N}{D(q_{\perp})D(q'_{\perp})}\Bigg[ \frac{4k_{\perp}^2 m^2}{x(x-1)(k_1\Delta_2-k_2\Delta_1)}\Bigg].
\ee \\

The relation of these GTMDs with that in \cite{metz} can be written as : 
 
\be
S^{0,-}_{1,ia} &=&2G_{1,4}^g;\nn \\
S^{0,-}_{1,ib} &=& -G_{1,1}^g;\nn \\
P^{0,-}_{1,ia} &=&  \frac{2m^2 G_{1,2}^g- \Delta_{\perp}^2 G_{1,1}^g}{2m^2};\nn \\
P^{0,-}_{1,ib} &=& \frac{2m^2G_{1,3}^g + k_{\perp}.\Delta_{\perp} G_{1,1}^g}{2m^2}.\nn \\
\label{re2}
\ee

\noindent
The spin-orbit correlation factor for the gluons $C_z^g$ can be defined in terms of the GTMD 
$G_{1,1}^g$ as follows \cite{other}, similar to quark case \cite{lorce_14} 
 \be
 C_z^g =  \int dx d^{2}k_{\perp} \frac{k_{\perp}^2}{m^2} G_{1,1}^g.
 \ee

\noindent
The GTMD $G_{1,1}^g$ calculated using Eq.~(\ref{re2}) agrees with that in \cite{other}
:
 
\be
 G_{1,1}^g = - \frac{N}{D(q_{\perp})D(q'_{\perp})}\Bigg[\frac{2(x^2-2x+2)m^2}{x^3(1-x)}   \Bigg];
\label{g11} 
 \ee

 \noindent
So the spin-orbit correlation factor for the gluons in the dressed quark model is
given by :
 
\be
C_z^g =    - N \int dx \frac{2(x^2-2x+2)(1-x)}{x}\Big[ I_{1} -
 m^2x^2 I_{2}\Big ].
\ee%

\noindent
The kinetic OAM for the gluons can be calculated using the sum rule for the gluon GPD's \cite{Ji} :
 
\be 
L^{g}_{z} = \frac{1}{2} \int dx \{ 
x [ H^{g}(x,0,0) + E^{g}(x,0,0) ] - \tilde{H^g}(x,0,0)
\}. \nn
\ee

\noindent
The gluon GPDs in the above relation  can be related to the GTMDs as follows:-
\begin{align}
  H^g(x,0,t) &= \int d^{2} k_{\perp} F_{1,1}^g;  \\
 E^g(x,0,t) &= \int d^{2} k_{\perp} \Big[ 
 -F_{1,1}^g +2\Big( 
 \frac{k_{\perp}.\Delta_{\perp}}{\Delta_{\perp}^{2}} F_{1,2}^g + F_{1,3}^g
 \Big)
 \Big];  \\
 \tilde{H}^g(x,0,t) &= \int d^{2} k_{\perp} G_{1,4}^g.
  \end{align} \\

 \noindent 
Using the above relation and the gluon GTMD calculated above, 
we can write the kinetic gluon OAM in the dressed quark model as
\be \label{cl}
 L^{g}_{z} = \frac{N}{2} \int dx \Big \{ 
-f(x) I_{1}  +  2\pi (1-x)
\Big
\};
\ee 
 where,
 
\begin{align}
  I_{1} &=   \int 
 \frac{d^{2} k_{\perp}}{m^2x^2 +(k_{\perp})^2}  = \pi log\Bigg[\frac{Q^2+m^2x^2}{\mu^2+m^2x^2}\Bigg];\nn \\
  f(x) &=2x^2-3x+2. 
\nn 
\end{align}
 
Unlike for the quarks \cite{our1}, canonical gluon OAM and and spin-orbit
correlations are different in this model. Note that the GTMDs $F_{1,4}$ and
$G_{1,1}$ depend on the gauge link. But up-to $O(\alpha_s)$ the result
does not depend on the choice of the gauge link \cite{other}.   
 
\begin{figure}[!htp]
\begin{minipage}[c]{1\textwidth}
\tiny{(a)}\includegraphics[width=8cm,height=7.5cm,clip]{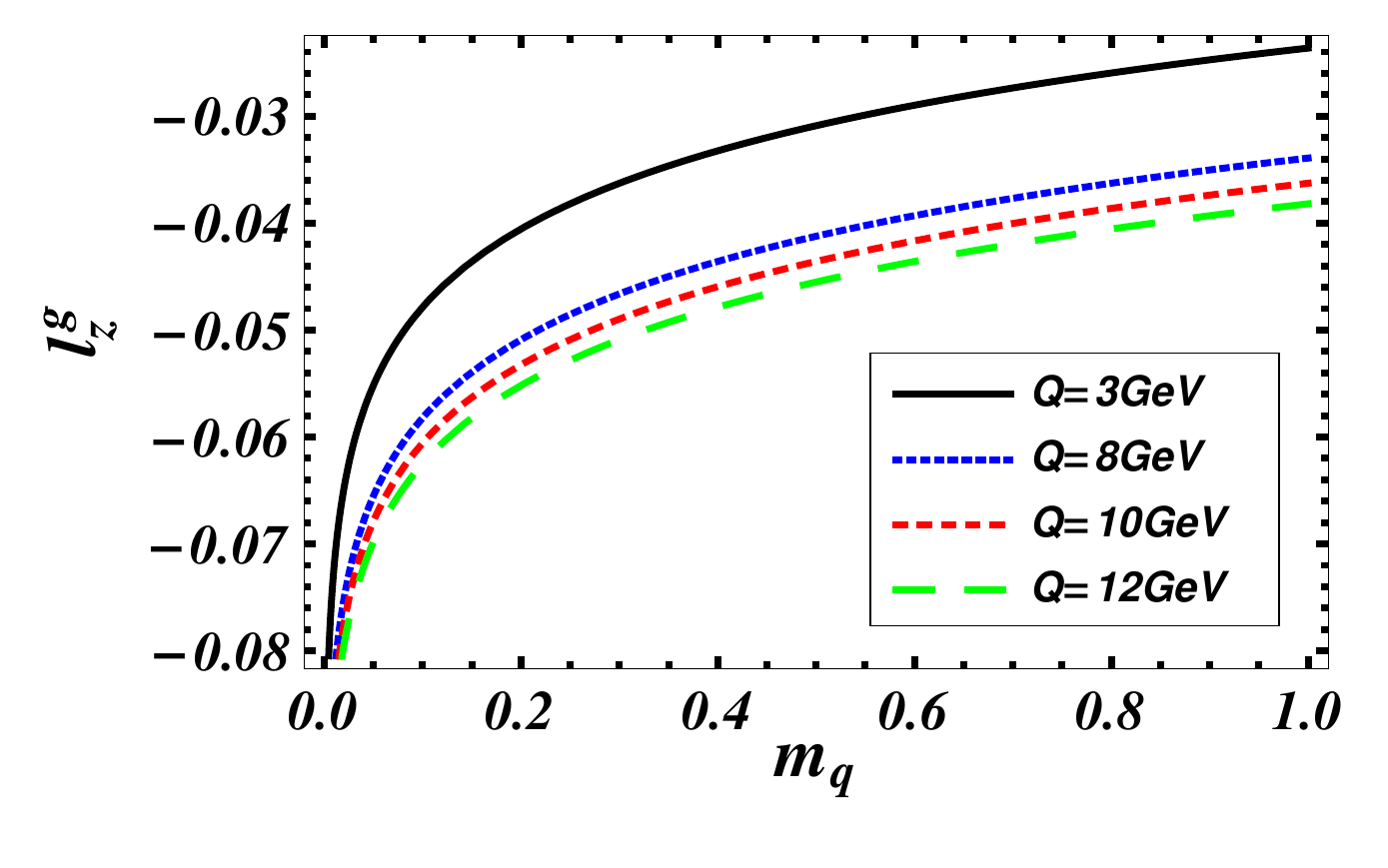}
\hspace{0.1cm}
\tiny{(b)}\includegraphics[width=8cm,height=7.5cm,clip]{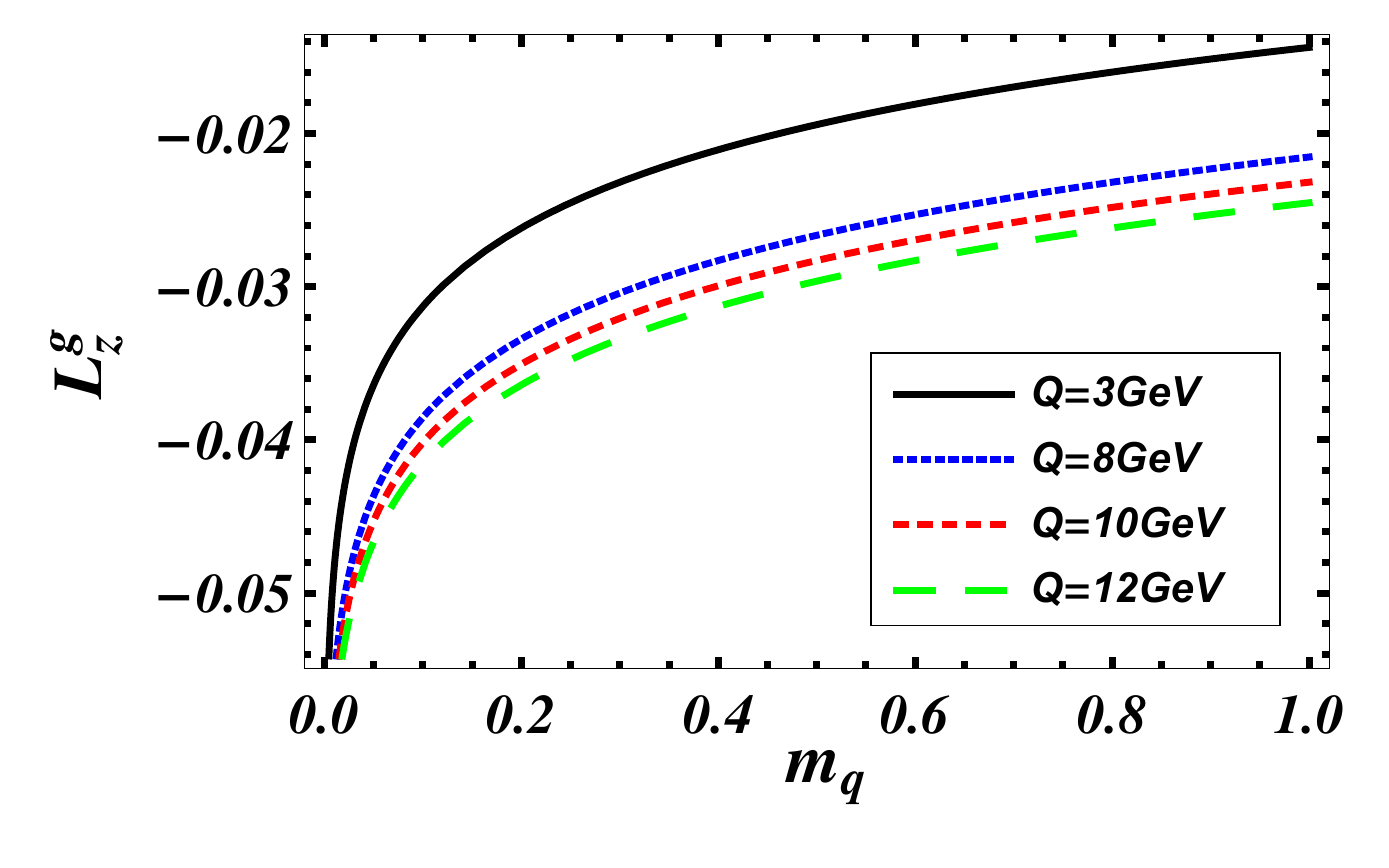}
\end{minipage}
\caption{\label{fig5}(Color online)
Plots of OAM (a) $l^{g}_z$ and (b) $L^{g}_z$ vs $m_q$  (GeV) for different values of $Q $
(GeV).
 }
\end{figure}

\section{Numerical Results}

\noindent 
In Figs.~\ref{fig1}~-~\ref{fig4}, we have shown the 3D plots for the Wigner
distributions of  gluon in the impact parameter space 
($b_x$-$b_y$), momentum  space ($k_x$-$k_y$) and also in the mixed space  ($k_y$-$b_x$). 
Normally the upper limit of the Fourier transform should be infinite. But in our
numerical calculation, we chose an upper limit of $\mid \Delta_\perp \mid $ which we
called $\Delta_{max}$.  Plots on the left and right column are for $\Delta_{max}= 1$ GeV
and $\Delta_{max}= 5$ GeV respectively. Dependence of gluon Wigner function on $\Delta_{max}$
is similar as the quark Wigner distributions: the peak of the Wigner distribution increases in 
magnitude as $\Delta_{max}$ increases. First, second and the third row in all  Figs.~\ref{fig1}~-~\ref{fig4} correspond to the 
impact parameter, momentum and mixed space plots respectively.
The plots in mixed space have probabilistic interpretation since we have integrated out the variable in the remaining direction 
i.e. $k_x$ and $b_y$ and for the impact parameter and momentum space plots 
these remaining variables are held constant. For all the plots, we have taken mass of 
target state to be 0.33 GeV. Also we integrated over $x$ and divided by the normalization constant $N$.

\noindent
In Fig.~\ref{fig1} , we show the 3D plots for the Wigner distribution of unpolarized gluon in an unpolarized target
state ($W^{UU}$). In Figs. 1(a) and 1(b) we see the variation of $W^{UU}$ in the position space. 
The magnitude of $W^{UU}$ is maximum at center ( $b_x=b_y=0$) and increases with increase in 
$\Delta_{max}$, which is expected from the analytic expression of $W^{UU}$. In Figs.  1(c) and 1(d) we
have plotted $W^{UU}$ in the momentum space for $\vec{b_\perp} = b \hat{j}=0.4$. In momentum space
too $W^{UU}$ peaks at center ($k_x=k_y=0$) and its magnitude increases with increasing 
$\Delta_{max}$. In Figs.  1(e) and 1(f) we have shown the variation of $W^{UU}$ in the mixed space. We
observed that $W^{UU}$ is maximum for $k_y=0$. As we move away from $k_y=0$, it first decreases
and then increases. Hence the probability to find a gluon in the target state is maximum near $k_y=0$.\\
It is worth mentioning here that similar plots for the quark Wigner
distributions in \cite{our1}  are rotated through an angle ${\pi \over 4}$ because
there we took $\Delta_\perp$ to be positive only whereas here we took
$\Delta_\perp$ to be both positive and negative.

\noindent
In Fig.~\ref{fig2}, we show the 3D plots for the Wigner distribution of unpolarized gluon in longitudinally polarized
target state. In Figs. 2(a) and 2(b) we see how $W^{LU}$ varies in position space. We
observed dipole structure whose magnitude increases with increase in $\Delta_{max}$. In
Figs. 
2(c) and 2(d) we have plotted $W^{LU}$ in the momentum space for a fixed 
$\vec{b_\perp} = b \hat{j}=0.4$. Again we observe a dipole structure but the polarity is flipped when compared to 
the plots in position space. Also, the magnitude of the peak increases
with increasing $\Delta_{max}$ which is expected from the analytic expression of 
$W^{LU}$. In Figs.  2(e) and 2(f) we have shown the variation of $W^{LU}$ in the mixed space. Here we
observe quadrupole structure whose magnitude increases with increase in $\Delta_{max}$.\\

\noindent
In Fig. ~\ref{fig3}, we show the 3D plots for the Wigner distribution showing the distortion 
due to the longitudinal polarization of the gluons. In Figs.  3(a) and 3(b) we show the
variation of $W^{UL}$ varies in position space for fixed $\vec{k_\perp} = k \hat{j}=0.4$.
We observe that the behavior is similar to the case of $W^{LU}$ showing a dipole structure. In
Figs.  3(c) and 3(d) we
have plotted $W^{UL}$ in the momentum space for $\vec{b_\perp} = b \hat{j}=0.4$. In the momentum space we observe a dipole 
like structure again similar to 
the case of $W^{LU}$ but here the polarity is not flipped unlike that in $W^{UL}$ when compared to the plots in the position space. In the mixed space we again observe a quadruple structure with increasing magnitude as $\Delta_{max}$ increases.

\noindent
In Fig. ~\ref{fig4}, we show the 3D plots for the Wigner distribution showing the distortion due to the correlation between the longitudinal polarization of the target state and the gluons.
In Figs.  4(a) and 4(b) we see how $W^{LL}$ varies in position space. In the b-space the behavior is similar to that shown by $W^{UU}$ and the magnitude increases with increasing $\Delta_{max}$ value.
In Figs.  4(c) and 4(d) we
have plotted $W^{LL}$ in the momentum space for $\vec{b_\perp} = b \hat{j}=0.4$. In momentum space, 
$W^{LL}$ shows a behavior similar to that of $W^{UU}$  and its magnitude increases with increasing 
$\Delta_{max}$. In the mixed space again the nature is identical to that shown by  $W^{UU}$.

\noindent
In Fig~\ref{fig5}, we have plotted the OAM of gluon with respect to mass of the target state for different
values of $Q$ where $Q$ and $\mu$ are the upper and lower limits of transverse momentum integration respectively.
 $\mu$ can be taken to be zero if the quark mass is non-zero.  In fact we
take taken $\mu$ to be zero. 
In Figs.  5(a) and 5(b) we show the canonical and the kinetic gluon OAM respectively as a function of the target mass.
We observe that the magnitude of both the OAM decreases with increasing mass of target state.

\section{Conclusion}
\noindent
In this work, we presented a calculation of gluon Wigner distributions for a
quark state dressed with a gluon. This can be thought of as a simple
composite spin ${1/2}$ system having a gluonic degree of freedom.  We showed
the various correlations between the  gluon spin and the spin of the 
target. We calculated the gluon kinetic and canonical OAM and also calculated the spin-orbit interaction of
the gluons. The kinetic and canonical OAM of the gluons differ in magnitude.
In most phenomenological models, there is no gluonic degree of freedom and a
study of gluonic contribution to the spin and OAM is not possible in such
models. Our simple field theoretical model calculations may be considered as
a first step towards understanding the gluon spin and OAM contribution.

\section{ACKNOWLEDGMENTS}
\noindent
We would like to thank C. Lorce for helpful discussions. This work is
supported by the DST project SR/S2/HEP-029/2010, Govt. of India.

\end{document}